%% file: paper.tex
\documentclass[useAMS,usenatbib]{mn2e}

\usepackage{graphicx}
\usepackage{times}
\usepackage{amsbsy}
\usepackage{amssymb}
\usepackage{amsmath}
\usepackage{natbib}
\usepackage[draft]{hyperref}

\newcommand{\bzero}{{\it B=0}}
\newcommand{\prim}{{\it primordial} }
\newcommand{\primR}{{\it primordial2R} }
\newcommand{\primRR}{{\it primordial3R} }
\newcommand{\agn}{{\it astrophysical} }
\newcommand{\agnR}{{\it astrophysicalR} }
\newcommand{\agnRR}{{\it astrophysical1R} }
\newcommand{\crpropa}{{\sc CRPropa} }
\newcommand{\ENZO}{{\sc ENZO} }

\newcommand{\isotropic}{{\it homogeneous} }
\newcommand{\LSS}{{\it density} }
\newcommand{\mass}{{\it mass halo} }

\newcommand{\bff}[1]{{#1}}
\newcommand{\bbff}[1]{{ #1}}

\usepackage{color}
\newcommand{\rbff}[1]{{ #1 }} 

\newlength{\onewidth}
\newlength{\minipagewidth}
\newlength{\skywidth}
\newlength{\skycolumn}
\newlength{\threewidth}
\begin{document}

\pubyear{2017}
\volume{e-print}
\label{firstpage}
\pagerange{\pageref{firstpage}--\pageref{lastpage}}
 
\title[UHECRs and origin of CMFs]{Simulations of ultra-high Energy Cosmic Rays in the local Universe and the origin of Cosmic Magnetic Fields}
\author[Hackstein et al.]{S. Hackstein$^{1}$\thanks{
 E-mail: shackste@physnet.uni-hamburg.de}, F. Vazza$^{1,2}$, M. Br{\"u}ggen$^{1}$, J. G. Sorce$^{3,4}$, S. Gottl{\"o}ber$^{4}$\\
$^{1}$ Hamburger Sternwarte, Gojenbergsweg 112, 21029 Hamburg, Germany\\
$^{2}$ INAF, Istituto di Radioastronomia di Bologna, via Gobetti 101, I-41029 Bologna, Italy \\
$^{3}$ Universit\'e de Strasbourg, CNRS, Observatoire astronomique de Strasbourg, UMR 7550, F-67000 Strasbourg, France \\
$^{4}$ Leibniz Institute for Astrophysics Potsdam, An der Sternwarte 16, 14482 Potsdam
 	}
\date{Accepted 2017 December 27. Received 2017 November 28; in original form 2017 August 6}
\maketitle

\begin{abstract}
	We simulate the propagation of cosmic rays at ultra-high energies, $\gtrsim 10^{18}$ eV, 
	in models of extragalactic magnetic fields in constrained simulations of the local Universe.
	We use constrained initial conditions with the cosmological magnetohydrodynamics code {\sc ENZO}.
	The resulting models of the distribution of magnetic fields in the local Universe are used in the \crpropa code to simulate the propagation of ultra-high energy cosmic rays.
	We investigate the impact of six different magneto-genesis scenarios, both primordial and astrophysical, on the propagation of cosmic rays over cosmological distances.
	Moreover, we study the influence of different source distributions around the Milky Way.
	Our study shows that different scenarios of magneto-genesis do not have a large impact on the anisotropy measurements of ultra-high energy cosmic rays.
	However, at high energies above the GZK-limit, there is anisotropy caused by the distribution of nearby sources,
	independent of the magnetic field model.
	This provides a chance to identify cosmic ray sources with future full-sky measurements and high number statistics at the highest energies.
\bbff{ 	Finally, we compare our results to the dipole signal measured by the Pierre Auger Observatory.
	All our source models and magnetic field models could reproduce the observed dipole amplitude with a pure iron injection composition.
	Our results indicate that the dipole is observed due to clustering of secondary nuclei in direction of nearby sources of heavy nuclei.
	A light injection composition is disfavoured by the non-observation of anisotropy at energies of $4-8 \rm\ EeV$.
}
\end{abstract}

\label{firstpage}

\section{Introduction}
\label{sec:intro}
\input{introduction.tex}
\section{Simulation}
\label{sec:simulation}
\input{simulation.tex}
\section{Results}
\label{sec:results}
\input{results.tex}
\section{Conclusions}
\label{sec:conclusions}
\input{conclusions.tex}
\section*{acknowledgments}
	Computations described in this work were performed using the \ENZO code (http://enzo-project.org), which is the product of a collaborative effort of scientists at many universities and national laboratories. 
	The constrained initial conditions have been developed in the context of the Constrained Local UniversE Simulations
(CLUES)  project (https://www.clues-project.org/).
	FV acknowledges financial support from the European Union's Horizon 2020 research and innovation programme under the Marie-Sklodowska-Curie grant agreement no. 664931 and the ERC STarting Grant MAGCOW, no. 714196.
	JS acknowledges support from the Astronomy ESFRI and Research Infrastructure Cluster ASTERICS project, funded by the European Commission under the Horizon 2020 Programme (GA 653477) .
	We acknowledge the usage of computational resources on the JURECA cluster at the at the Juelich Supercomputing Centre (JSC), under projects no. 10755, 11823, 9016 and 8998, and  on the Piz-Daint supercluster at CSCS-ETHZ (Lugano, Switzerland) under project s701.
	SH, FV and MB additionally thank G. Sigl and A. Dundovic for fruitful discussion as well as the other developers of \crpropa for making the code public.
	We further acknowledge the use of computational resources at the Rechenzentrum of the University of Hamburg.
	The whole group would also like to thank the referee for fruitful comments and careful reading of the manuscript.

\bibliographystyle{mnras}
\bibliography{cites}

\label{lastpage}

\appendix

\input{appendix}

\end{document}

%% file: introduction.tex
	Evidence for the existence of magnetic fields have been reported for all types of structures found throughout the Universe.
	Galaxies host magnetic fields with typical strengths of $\sim 5-15 \rm\ \mu G$, which were measured using Faraday rotation and synchrotron emission up to redshift $z\sim 2-6$ \citep[e. g. ][]{galactic_fields,2013ApJ...772L..28B,2016A&ARv..24....4B,2016ApJ...829..133K}.
	The magnetic field in clusters of galaxies was found to be of the order $\sim \rm \mu G$ \citep{Feretti2012}.
	Future radio observations will offer the chance to measure the magnetisation at the outskirts of clusters and in filaments that connect them \citep{2011JApA...32..577B,2012MNRAS.423.2325A,Vazza2015}.
	A recent study has reported upper limits on the magnetic field strength of $\sim 0.03 \rm\ \mu G$ from the absence of a correlation between synchrotron emission and the large-scale sructure \citep[LSS,][]{2017MNRAS.467.4914V,2017MNRAS.468.4246B}.
	Limits on the magnetic fields in voids were derived from the angular power spectrum, the bi- and the trispectrum of the cosmic microwave background \citep[$B_{\rm void} < 1 \rm\ nG$, ][]{PLANCK2015,2014PhRvD..89d3523T},
\bff{	absence of evolution with redshift in Faraday rotation measures \citep[$B_{\rm void} < 1.7 \rm\ nG$, ][]{2016PhRvL.116s1302P} }
	and the lack of secondary emission around blazar sources \citep[$B_{\rm void} > 10^{-7} \rm\ nG$, ][]{NeronovVovk2010,AlvesBatista:2017vob}\footnote{See however discussion in \citet{2012ApJ...752...22B}
	for a different view of the issue.}.
\\ \\ \\
	Magneto-hydrodynamical (MHD) cosmological simulations have been used to evolve magnetic fields of primordial or other origin that are amplified during structure formation and by additional dynamo processes \citep[e. g.][]{2006AN....327..575D,2012SSRv..166....1R}.
	These simulations produce models of cosmic magnetic fields (CMFs) that agree to some extent with observations \citep[e. g.][]{1999dtrp.conf..237D,2005ApJ...631L..21B,Donnert2008}.
	All amplification scenarios have in common that they require a seed field, whose structure, strength and origin is unknown.
\\ \\
	In this paper we probe the possibility to learn about the origin of CMFs using measurements of cosmic rays at ultra-high energies.
	Previous studies on similar topics mainly focussed on properties of the galactic magnetic field \citep{0004-637X-479-1-290,Takami:2007kq}
	or small-scale anisotropies \citep{2002JHEP...07..006H,Yoshiguchi:2002rb}.
	Other works used unconstrained MHD models to study the implications of CMFs on Ultra-high energy cosmic rays (UHECRs) astronomy \citep{Sigl:2003ay,2004PhRvD..70d3007S,Sigl2004,Kotera:2007ca,Das:2008vb,Hackstein2016},
	An overview of UHECR studies using MHD simulations can be found in \citet{AlvesBatista:2017vob}.
	Analytical studies on the implication of CMFs on UHECR observations are provided in \citet{Harari:2000az,2002JHEP...07..006H,2002JHEP...03..045H,Tinyakov:2004pw,Takami2012767}.
\\ \\
	In previous work \citep{Hackstein2016} we found strong variance in the observables of UHECRs induced by the position of, both, observer and sources \citep[also cf. e. g.][]{Sigl2004}.
	To reduce this cosmic variance it is necessary to use constrained MHD models that resemble the local Universe, as has been done by \citet{Dolag:2003ra}.
	They conclude that UHECR protons are reasonably deflected only when they cross galaxy clusters, though they assumed a rather weak field in voids of $\lesssim 10^{-11}$ G.
	Our new work expands the early work by \citet{Dolag:2003ra} in a few ways:
	a) we use the most recent set of initial conditions by \citet{2016MNRAS.455.2078S}, which were derived with more updated algorithms and observational constraints (see Sec. \ref{sec:initial_conditions}); 
	b) we relied on a different numerical method: i.e. the grid-MHD simulations with {\ENZO} instead of smoothed-particle hydrodynamics (SPH) simulations,
	which gives us a better sampling of moderate and low resolution regions; 
	c) we performed a survey of magnetic field models, rather than assuming a single specific scenario. 
\\ \\
	Data suggests that cosmic rays are fully ionized nuclei that constantly hit the Earth from outer space with energies that range over 11 orders of magnitude.
	At low energies ($< 10^{17}$ eV) the predominant sources were found to be super-nova remnants in our own Galaxy,
	where charged particles experience Fermi acceleration in magnetic shocks \citep[e. g. ][]{blasi}.
	UHECRs are less prone to the deflection in CMFs, thus they are not confined within their host galaxy and presumably are of extragalactic origin.
	The sources of UHECRs are currently unknown.
	If we assume the same acceleration process as at low energies, the size of the source limits the maximum energy of emitted UHECRs.
	This is the famous Hillas criterion \citep{hillas} that limits the candidates for sources of UHECRs at $\gtrsim 10^{20}$ eV to very few objects, namely radio galaxy lobes, clusters of galaxies, active galactic nuclei and gamma ray bursts \citep[e. g.][]{2016arXiv160407584D}.
	Recent works have reported signs of anisotropy in simulations with pure proton composition and limited source density in correlation with the LSS  \citep{2017arXiv170602534D, Abreu:2013kif}, that are not observed in nature.
	They infer lower bounds on the density of sources of $\sim 10^{-4} \rm\ Mpc^{-3}$.
	Also they conclude that the UHECR flux cannot be dominated by protons.
	In this work we investigate the effect of different source distributions of UHECRs on the observed arrival directions.
\\ \\
	This article is organized as follows: 
	In Sec. \ref{sec:simulation} we present details on the simulation of the MHD models and of the propagation of UHECRs.
	The results of these simulations are then discussed in Sec. \ref{sec:results}.
	Our conclusions are finally given in Sec. \ref{sec:conclusions}.

%% file: simulation.tex
\label{sec:simulation}
\subsection{Constrained initial conditions}
\label{sec:initial_conditions}
	Simulations that resemble the local Universe stem from particular initial conditions.
	 Unlike typical initial conditions that abide solely by a cosmological prior, 
	 these initial conditions are additionally constrained by local observational data that can be either redshift surveys  \citep{2010MNRAS.406.1007L,2013MNRAS.435.2065H} or radial peculiar velocities of galaxies  \citep{2002ApJ...571..563K,2003ApJ...596...19K, 2014MNRAS.437.3586S}. 
	We use the latter with a backward \citep[by opposition to forward][]{2013MNRAS.429L..84K,2013MNRAS.435.2065H,2013MNRAS.432..894J, 2014ApJ...794...94W} technique \citep{1987ApJ...323L.103B,1991ApJ...380L...5H,1992ApJ...384..448H,1993ApJ...415L...5G,1996MNRAS.281...84V,1998ApJ...492..439B,2008MNRAS.383.1292L}.
	The catalog of constraints is fully described in \citet{2013AJ....146...86T} and the method to produce the constrained initial conditions is summarized in \citet{2016MNRAS.455.2078S}.
	The process involves various steps from the minimization of biases \citep{2015MNRAS.450.2644S} in the catalog of peculiar velocities to the constrained realization technique \citep{1991ApJ...380L...5H} to get the final product: the initial conditions. 
	We work within the Planck cosmology framework \citep[$\Omega_m$=0.307, $\Omega_\Lambda$=0.693, h=0.677, $\sigma_8$~=~0.829,][]{2014A&A...571A..16P}.
\subsection{MHD-simulations}
\begin{table*}
\begin{tabular}{c|c|c}
	mnemonic & gas physics & magnetic field  \\ \hline
	\bzero & non-radiative & $B_0=0$ \\ \hline
	\prim & non-radiative &  $B_0=0.1$ nG  \\ \hline
	\primR & non-radiative & $(\langle B^2 \rangle)^{0.5} = 1$ nG, $n_B = -3$ \\ \hline
	\primRR &non-radiative & $(\langle B^2 \rangle)^{0.5} = 1$ nG, $n_B = -4$  \\ \hline
	\agn & cooling and AGN feedback & $5 \cdot 10^{58}$ erg, $z<4$; $B_0=10^{-11}$ nG  \\ \hline
	\agnR & cooling and AGN feedback & $10^{60}$ erg, $z<4$; $B_0=10^{-11}$ nG \\ \hline
	\agnRR & cooling and AGN feedback & $10^{60}$ erg to $5 \cdot 10^{58}$ erg, $z<1$; $B_0=10^{-11}$ nG
\end{tabular}
\caption{ List of magnetic field models investigated in this paper. 
	First column: name of the model;
	second column: physical module for the gas component;
	third column: generation of magnetic field.
	All models were simulated within a volume of $(500 \text{ Mpc}/h)^3$.
	In \crpropa we used the innermost $(250 \text{  Mpc}/h)^3$ with $1024^3$ data cells and a resolution of $245 \text{ kpc}/h$.}
\label{tab:model}
\end{table*}
\begin{table*}
\begin{tabular}{c|c|c|c|c}
	mnemonic & injection scenario & $L_{\rm\ box}$ & $N_{\rm\ sources}$ & $n_{\rm\ sources}$ \\ \hline
	& & [Mpc$/h$] & & [Mpc$^{-3}h^3$] \\ \hline
	\isotropic & random positions  & 250 & $10^8$ & 6,4 \\ \hline
	\LSS & same as \isotropic with $p.d.f. = \rho_{\rm gas} \ / \ \sum \rho_{\rm gas} $ & 250 & $10^8$ & 6,4 \\ \hline
	\mass & virial haloes, uniform luminosity & 250 & 2672 & $1.71\cdot 10^{-4}$
\end{tabular}
\caption{List of the injection models.
	First column: name of model; 
	second column: set of sources;
	third column: box length of the simulated volume; 
	fourth column: number of sources in the simulated volume; 
	fifth column: number density of sources.}
\label{tab:source}
\end{table*}
\begin{figure*}
\includegraphics[width=0.88\textwidth]{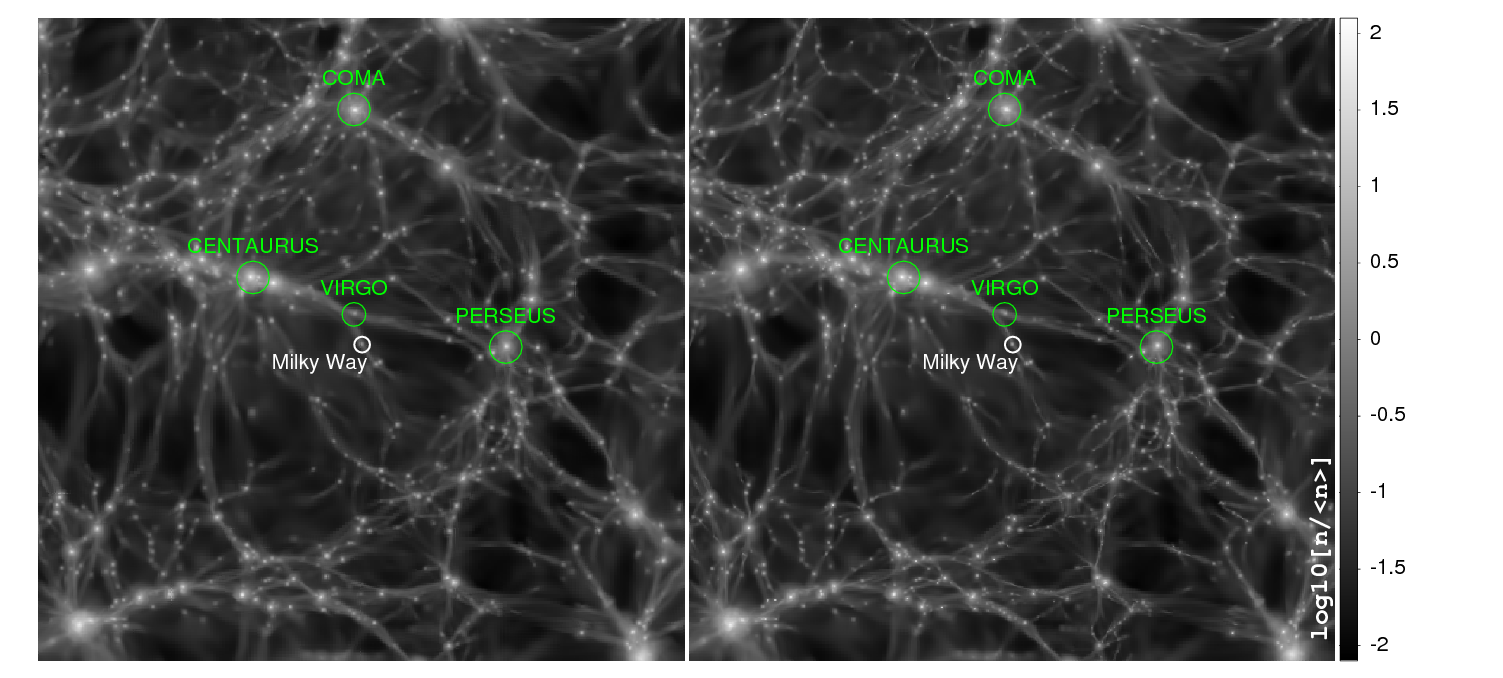}
\includegraphics[width=0.88\textwidth]{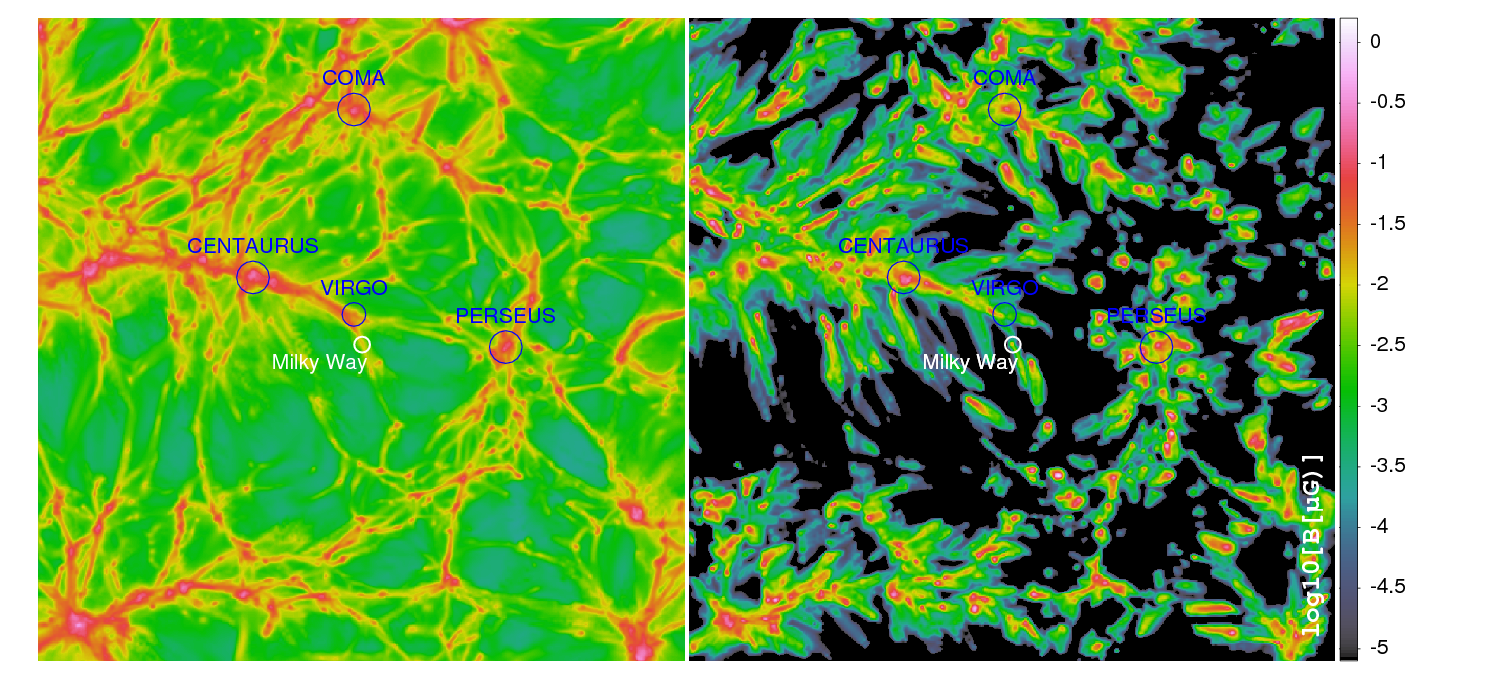}
\includegraphics[width=0.9\textwidth]{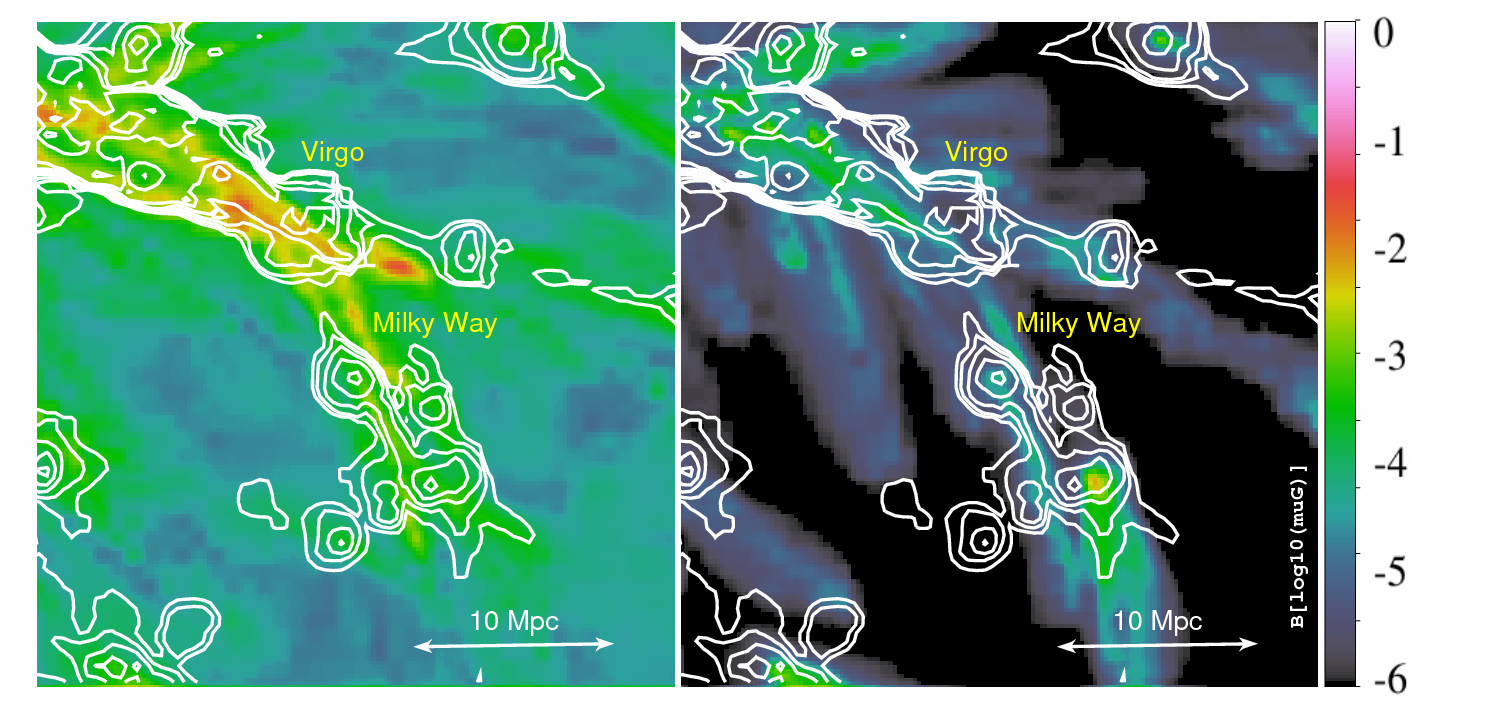}
  \caption{
	Maps of projected gas number density (top) and mean magnetic field along the line of sight (centre, bottom) for the \primR model (left) and for the \agnR model (right) at $z=0$.
\bff{	The gas number density $n$ is normalized to the average density in the whole volume, $n/<n>$.
	The magnetic field is shown in $\mu$G.
	Colours are in logarithmic scale. }
	The top and centre panels have a side-length of $200 \text{ Mpc}/h$, the projection axes are the $X$ and $Y$ in the super-galactic coordinates.
\bff{	The bottom panels give a more detailed view on the central $40 \text{ Mpc}/h$. }
	The position of the Milky Way observer considered in this work is exactly at the centre of the box, indicated by a white circle in the top and center panels.
	The additional circles show the location of the simulated counterparts of real objects in the local Universe.
 }
\label{fig:simulation} 
\end{figure*}
	The MHD simulations performed in this paper have been produced with the cosmological grid code \ENZO 
	that follows the dynamics of dark matter with a particle-mesh N-body method and uses a variety of shock-capturing Riemann solvers to evolve the gas component \citep{ENZO}.
	The MHD equations were solved with the method by \citet{Dedner2002}.
	To keep $\nabla \cdot \boldsymbol{B}$ as low as possible, it uses hyperbolic divergence cleaning.
	The fluxes at cell interfaces are reconstructed with the Piecewise Linear Method.
	They are evolved using the local Lax-Friedrichs Riemann solver \citep{KurganovTadmor2000},
	with time integration using the total variation diminishing  second order Runge-Kutta  scheme \citep{ShuOsher1988}.
	The set of simulations were run on Piz-Daint (CSCS) and made use of the recent implementation of the Dedner algorithm using CUDA \citep{2010NewA...15..581W}.
\\ \\
	To model the local Universe at $ z = 0 $, the MHD simulations started at $ z = 60 $ with initial conditions described in Sec. \ref{sec:initial_conditions}.
	We sampled a volume of $(500 \text{ Mpc}/h)^3$, with $512^3$ cells and dark matter particles.
	We use this large volume in order to remove effects from periodic boundary conditions in the constrained subregion of the MHD simulation.
	Since the initial perturbation for  baryonic matter are not provided in the initial conditions of Sec. \ref{sec:initial_conditions}, 
	we simply initialise baryons to the uniform cosmological density, assuming an initial zero-velocity field for baryons everywhere. 
	While more accurate ways to couple baryons to dark matter perturbations since the beginning are possible,
	this choice is irrelevant for the level of details we are concerned here \citep[e. g.][]{2011MNRAS.418..960V}.
	Full resolution of the whole box is not necessary and costly,
	therefore only the constrained innermost $(200 \text{ Mpc}/h)^3$ volume was further refined by a factor 32 using adaptive-mesh refinement.
	The refinement here follows the standard local over-density criterion, doubling the cell resolution whenever the local gas over-density was 3 times larger than the surroundings, up to a maximum of 5 levels of refinement ($\approx 31 \text{ kpc}/h$ per cell). 
	The clusters that form in this volume closely resemble real local structures (e. g. the Centaurus, Virgo, Coma and Perseus clusters),
	within typical offsets of order $\leq 2-3 ~\text{ Mpc}/h$ which are however not crucial for the global studies we perform here. 
\\ \\
	The limited size of computer memory used for the simulation of UHECR propagation did not allow to use the full volume of the CMF models obtained from MHD simulations.
	In order to minimize effects from periodic boundaries (see Sec. \ref{sec:crpropa}),
	we restricted the simulations in \crpropa to the innermost $(250 \text{ Mpc}/h)^3$ volume and reduced the number of cells inside that volume to $1024^3$.
	The resulting resolution is then $245 \text{ kpc}/h$. 
\\ \\
	The use of constrained simulations of the local Universe is an important step forward compared to our previous work \citep{Hackstein2016},
	where we found a large variance in the observed properties of UHECRs from observer to observer. 
	Given the strong impact of $\leq 35$  Mpc$/h$ sources of UHECRs, 
	it is not granted that the average over many observers is representative of what can be observed by the specific observer at Earth's location.
	However, in these new runs placing our observer within the Local Group allows us to remove these uncertainties. 
	At distances $> 100$ Mpc$/h$, deflection and the increasing number of sources provide an UHECR flux almost independent of the exact position of distant sources.
	It is therefore sufficient to model the source distribution only within that distance.
\\ \\
	Following a procedure similar to \citet{Hackstein2016}, we ran several MHD simulations with different scenarios for the origin of CMFs.
	In the \prim model we used a uniform initial magnetic field of strength 0.1 nG (comoving) along each axis at $z = 60$.
	In the \primR and \primRR models, similar fields were generated by drawing the magnetic field from an analytically generated power-law distribution of magnetic fields,
	with two different slopes for the power spectrum, $n_B= -3$ and $n_B=-4$, respectively (with $P_B \propto k^{n_B}$), see \citet{2016A&A...594A..19P} for details.  
	We have generated a power-law spectrum distribution of the vector potential in the Fourier space for a $1024^3$ grid, randomly drawn from the Rayleigh distribution,
	and we have computed the magnetic field in real space as $\vec{B} = \nabla\times A$, ensuring $\nabla\cdot \vec{B} = 0$ by construction. 
	We have assumed that the maximum coherence scale of the magnetic field is $500 \text{ Mpc}/h$ and that the minimum scale is the root grid resolution,
	and that the power-law of fluctuations follows the input $P_B$ power spectrum, similar to \citet{Bonafede2013}. 
	In both cases, the normalisation of the spectrum of initial fluctuations is chosen such that $(\langle B^2 \rangle)^{0.5}=B_0$, i.e. the rms magnetic field is equivalent to the uniform seeding case. 
\\ \\
	The astrophysical origin of CMFs was modelled as impulsive thermal and magnetic feedback in haloes
	where the physical gas number density exceeded a critical value of $10^{-2} \rm\ cm^{-3}$.
	The thermal energy is released as a couple of over-pressurized outflows at random opposite directions from the halo centre.
	The feedback magnetic energy, assumed to be 50\% of the injected thermal energy, is released as dipoles around the centre.
\\
	In the \agn model, we assumed a release of $5 \cdot 10^{58}$ erg per feedback episode starting from $z=4$; 
	in the \agnR model we used instead a larger budget of $10^{60}$ erg per event. 
	Finally, in the \agnRR model we considered a mixed scenario, where we changed the energy budget from  $10^{60}$ to $5 \cdot 10^{58}$ erg per event  from  $z=1$ to $z=0$.
\\
	All runs with astrophysical scenarios for the emergence of extragalactic magnetic fields used equilibrium radiative gas cooling, assuming a fixed metallicity of $Z=0.3 Z_{\odot}$. 
	While the cooling is necessary to trigger the onset of cooling flows and start the cooling-feedback cycle in our halos, the large-scale distribution of gas matter outside simulated halos is similar across all runs (see Sec. \ref{sec:magnetic_fields}). 
\\
	In all astrophysical runs we impose a uniform lower magnetic field level of $B_0=10^{-20} \rm G$ comoving at $z=60$. 
	This extremely low magnetisation prevents the formation of spurious numerical effects at the boundary between magnetised and unmagnetised regions in the simulation (in contrast to the primordial models, where there is a non-zero magnetic field everywhere). 
	An overview of the models is given in Tab. \ref{tab:model}.
\subsection{UHECR simulations}
\label{sec:crpropa}
	The resulting CMF models \bff{ used in \crpropa have a volume of $(250 \rm\ Mpc/h)^3$, discretised by $1024^3$ cells of $(244 \rm\ kpc/h)^3$ volume that contain a uniform field.
	These models} are used to simulate the propagation of UHECRs in the local Universe in order to search for different signatures in the UHECR arrival directions.
	This is done with \crpropa3.0\footnote{\href{https://crpropa.desy.de}{https://crpropa.desy.de}} \citep{CRPropa2006,CRPropa2013,CRPropa2016},
	a publicly available code to study the propagation of UHECRs.
	\crpropa computes all the relevant processes of propagation,
	this includes Lorentz deflection, energy loss by production of particles and cosmic expansion, photo disintegration and nuclear decay.
	The code further allows to track the trajectories of particles in a 3D volume.
\\ \\
	We let \crpropa inject $10^8$ protons with random momentum from random positions.
	\bff{ The initial energies range from 1 EeV to $10^3$ EeV, following a power spectrum of $E^{-1}$.
	This choice does not result in the energy spectrum observed in nature, but was used in order to increase number statistics at the highest energies.\footnote{\rbff{A steeper injection spectrum would result in too low accuracy of the measurement of anisotropy around 100 EeV, as can be seen by Eq. 1.}}
\\
	After injection, the energy loss and trajectories of the particles are calculated. }
	In case a trajectory leaves the volume, it is continued on the opposite side.
	An event is recorded when a trajectory intersects with the observer.
	This observer is represented by a sphere of radius 800 kpc in the centre of the simulation, which is the defined position of the observer in a constrained simulation.
	For a discussion on the role of the finite observer size in \crpropa simulations we refer the reader to \citet{Hackstein2016}.
\\
\bff{ 	After intersection, trajectories continue so they may reach another replica of the observer.
	Environments with strong magnetic fields can trap particles so they arrive at the same observer again.
	If the same particle is recorded multiple times at the same observer, we randomly chose one of these events.
	This choice excludes over-counting of trapped particles and no further weighting is necessary.
}
\\
	In a different set of runs we repeat the process with $10^7$ iron nuclei,
	taking care also of nuclear decay and disintegration processes and follow the trajectories of secondary nuclei.
\\ \\
	In order to investigate the influence of the distribution of sources, we tested different source models for UHECRs  in all the CMF models listed above.
	In order to bracket the present uncertainties on the degree of isotropy in the distribution of sources,
	we analyse the extreme case of a \isotropic model, in which we inject each particle at a random position anywhere in the simulated volume.
\bff{ 	This mimics the absence of structure in the distribution of sources and shows the impact of source distribution in comparison to the other models. }
\\
	It is generally assumed that sources of UHECRs are powerful sources located in galaxies.
	Therefore we assume that the distribution of sources correlates with the LSS. 
	In the \LSS model, particles are injected at random positions with a probability density function identical to the gas density, renormalized by the total gas density in the volume, $p.d.f. = \rho_{\rm gas} / \sum \rho_{\rm gas}  $.
\bff{ 	This model with maximum source density reflects a huge number of transient sources that may be found in all types of galaxies, such as gamma-ray bursts or magnetars. }
\\
	Finally, the \mass model agrees with the lower bounds on source density \citep[$\sim 10^{-4} \rm\ Mpc^{-3}$, ][]{Abreu:2013kif}, 
	where we take as sources the centres of 2672 virial haloes identified in our simulation, each with the same luminosity of UHECRs.
\bff{ 	This model mimics the case of very few stationary sources, e. g. radio galaxies or AGN.
	The precision of the MHD-simulations did not allow to resolve these structures individually. }
	An overview of the source models can be found in Tab. \ref{tab:source}.
\vspace{-12pt}

%% file: results.tex
\subsection{Simulated extragalactic magnetic fields}
\label{sec:magnetic_fields}
	Figure \ref{fig:simulation} shows the maps of projected gas density (top) and of mean magnetic field along the line of sight (centre,  bottom) for the \primR run (left) and for the \agnR run (right) at $z=0$.  
	While the different implementations for gas physics do not significantly change the distribution of gas matter on large scales,
	the differences in the assumed magneto-genesis scenarios affect the morphological distribution and strength of extragalactic magnetic fields.
\\ \\
\begin{figure}
\includegraphics[width=0.5\textwidth]{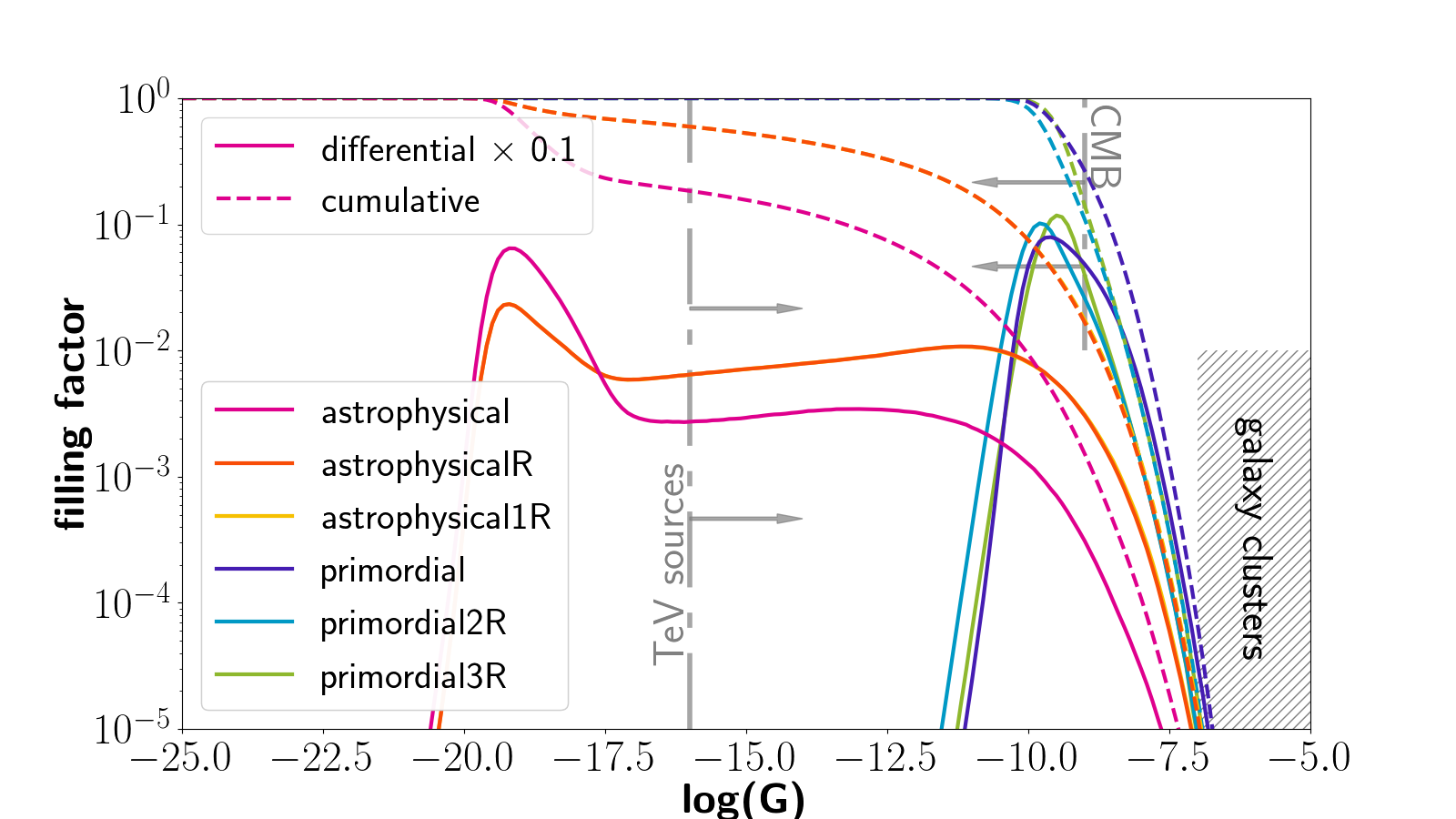}
\caption{Volume filling factor of the models listed in Tab. \ref{tab:model}.
	The solid lines show the differential filling factor renormalized by 0.1 for clarity, dashed lines show the cumulative filling factor.
	The grey arrows and shaded area indicate the limits given from observations as listed in the introduction.
	The yellow line of the \agnRR model fits exactly with the \agnR model.
	}
\label{fig:filling_factor}
\end{figure}
	In Fig. \ref{fig:filling_factor}, we present the volume filling factor of the models listed in Tab. \ref{tab:model}.
	All models have magnetic fields in cluster regions that agree with observational limits.
	The different \prim models show very similar filling factors with dominant strength at $\sim 0.1 \rm\ nG$, 
	close to the upper limit on magnetic field strength in voids from analysis of the CMB anisotropy \citep{2014PhRvD..89d3523T,PLANCK2015}.
\\
	The strong fields in the \agn models are concentrated in the dense regions of the simulation,
	which are predominantly filled with very weak fields, at odds with lower limits inferred from the lack of secondary emission around blazar sources \citep{NeronovVovk2010}.
	The filling factors of the \agnR and \agnRR models are almost identical, only in the \agn model an even smaller volume contains strong fields.
	Due to the later seeding of magnetic field in all of the astrophysical models, as compared to the primordial models,
\bff{ 	more of the original, oriented field components survive until $z = 0$ } and thus a greater influence on the propagation of UHECRs is expected.

\subsection{Energy spectrum}
\label{sec:energy_spectrum}
\begin{figure*}
\includegraphics[width=0.495\textwidth]{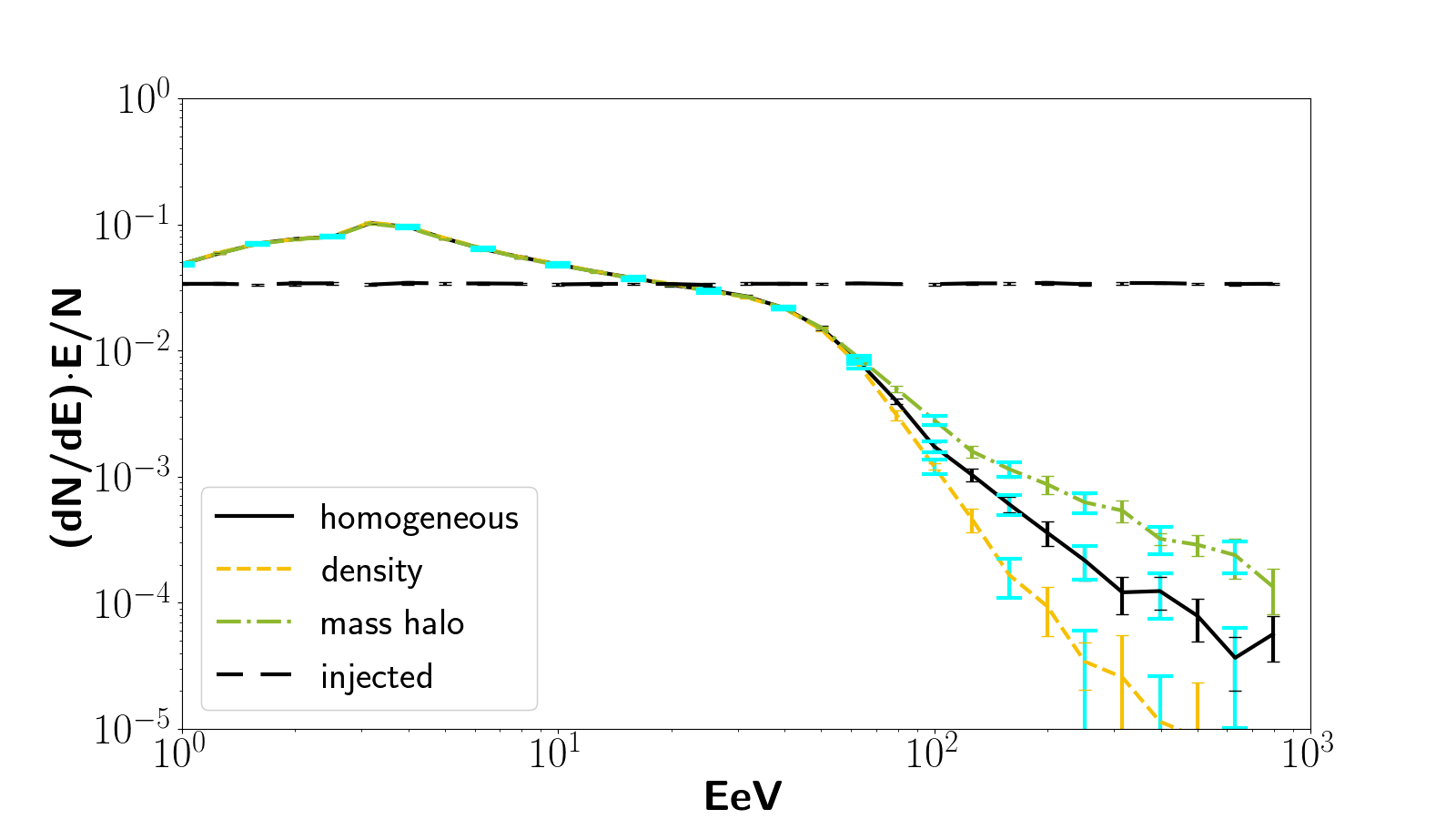}
\includegraphics[width=0.495\textwidth]{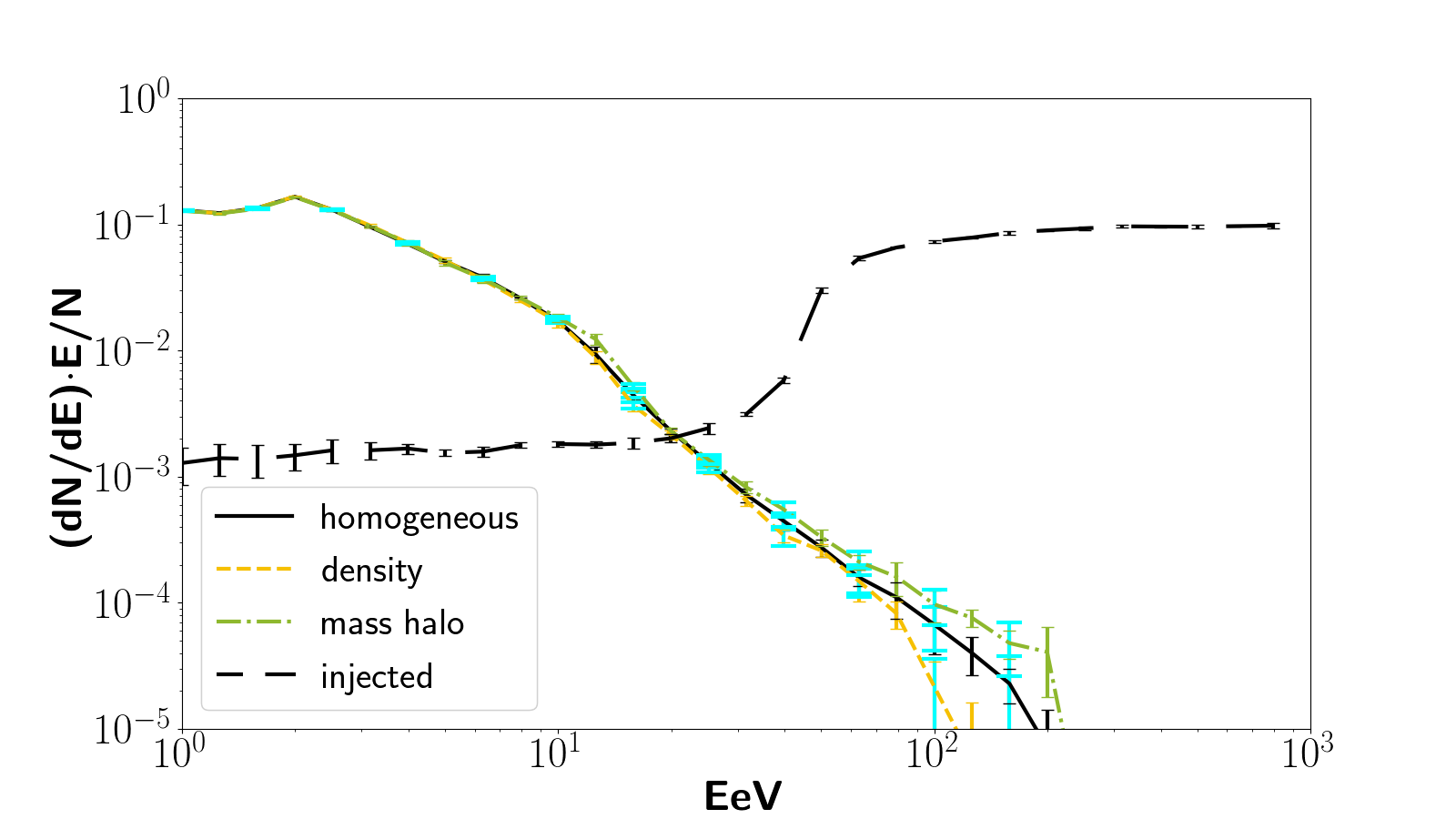}
\caption{Energy spectrum of UHECRs as injected at the sources (dashed lines) and measured by the observer  for a pure proton and a pure iron injected composition (left and right, respectively).
	The colours and line styles indicate the injection models listed in Tab. \ref{tab:source}.	
\bbff{The graphs show the average over all magnetic field models, the standard deviation is indicated by the narrow error bars.
	The big cyan error bars show the Poisson noise at each second data point.}
	For clarity, the graphs are multiplied by the inverted energy spectrum at injection $E$,
	and renormalized with the total number of particles $N$.
	}
\label{fig:energy_spectrum}
\end{figure*}
	In Fig. \ref{fig:energy_spectrum}, we show the energy spectrum of UHECRs as injected at the sources and measured by the observer.
	For clarity, the graphs are renormalized by the total number of observed events $N$ and
\bbff{multiplied by the inverted energy spectrum at injection, which was set to be $E^{-1}$. } 
	Below 100 EeV, the energy spectrum is universal, as predicted by the propagation theorem \citep{2004ApJ...612..900A}.
	In particular, we find no influence of the underlying magnetic field on the observed energy spectrum, as has been shown in \citet{Hackstein2016}. 
\\
\bbff{In the proton injection scenarios the total number of observed events is $N = 50,000$ with 15,000 events above 10 EeV.
	In the iron injection scenarios, $N=100,000$ with 5,000 events above 10 EeV.
	The fluctuation of these numbers between scenarios with the same initial composition is about 10\%. 
	Therefore, number statistics of the observables presented in Sections \ref{sec:APS} \& \ref{sec:composition} are comparable.
}
\\ \\
	In the proton injection scenarios, the slope above the GZK cut-off is not universal but shows significant variation in different source models.
\bbff{We show, both, the Poisson shot-noise as well as the standard deviation for different magnetic field models.
	They are almost identical, i. e. the error is dominated by statistical fluctuations.
	Magnetic fields leave no significant impact, as expected for quasi-rectilinear propagation.
}\\
	The spectrum at $\sim 100$ EeV is significantly harder in the \mass injection model, 
	where there is an above-average amount of sources within a few Mpc of the observer.
	Furthermore, in the \isotropic and \mass injection models protons with up to 800 EeV arrive at the observer in all magnetic field models.
	However, in the \LSS injection senario there are no particles received above 400 EeV.
	The closer the nearby sources, the higher the number of events that are observed at the most extreme energies and the higher the maximum energy of observed events.
\\ \\
\bbff{Most particles injected in the simulation never reach the observer and are lost.
	The injected spectrum plotted in Fig. \ref{fig:energy_spectrum} only shows the injected energies of particles received by the observer.
	In the proton case the injected spectrum of observed particles perfectly recreates the injection spectrum used for simulation.
}
	In the iron case, multiple secondary nuclei of the same nucleus can reach the observer.
\bbff{In the injected energy spectrum the primary nucleus is counted once for every secondary nucleus that is observed.}
	This double counting accounts for the sharp increase in the injected spectrum above 40 EeV.
	At low energies,   $\gtrsim 1$ EeV, the injected spectrum is slightly decreased in the stronger magnetic field models. 
	Iron nuclei at low energy are deflected more strongly and are more likely to lose their energy before they reach the observer.
\\
	The slope of the observed spectrum is much steeper than in the proton case.
	The low energies are dominated by the secondary protons of iron injected at the highest energies.
	Only few events are observed with energies $>$100 EeV.
	This is because most of heavy nuclei at those energies disintegrate completely within a few Mpc and distribute their energy evenly among their secondary protons \citep{1998JHEP...10..009E,2012APh....39...33A}.
	Thus, too few events are observed in the iron injection case to measure deviation from isotropy.
\\
	In conclusion, a sharp cut-off, as observed by extensive air shower arrays \citep{2010ApJ...712..746I,2014BrJPh..44..560L},
	would hint at a low number of nearby sources or a maximum acceleration energy of protons at the sources that is below the cut-off.
\subsection{Angular power spectrum}
\label{sec:APS}
\begin{figure}
\includegraphics[width=0.5\textwidth]{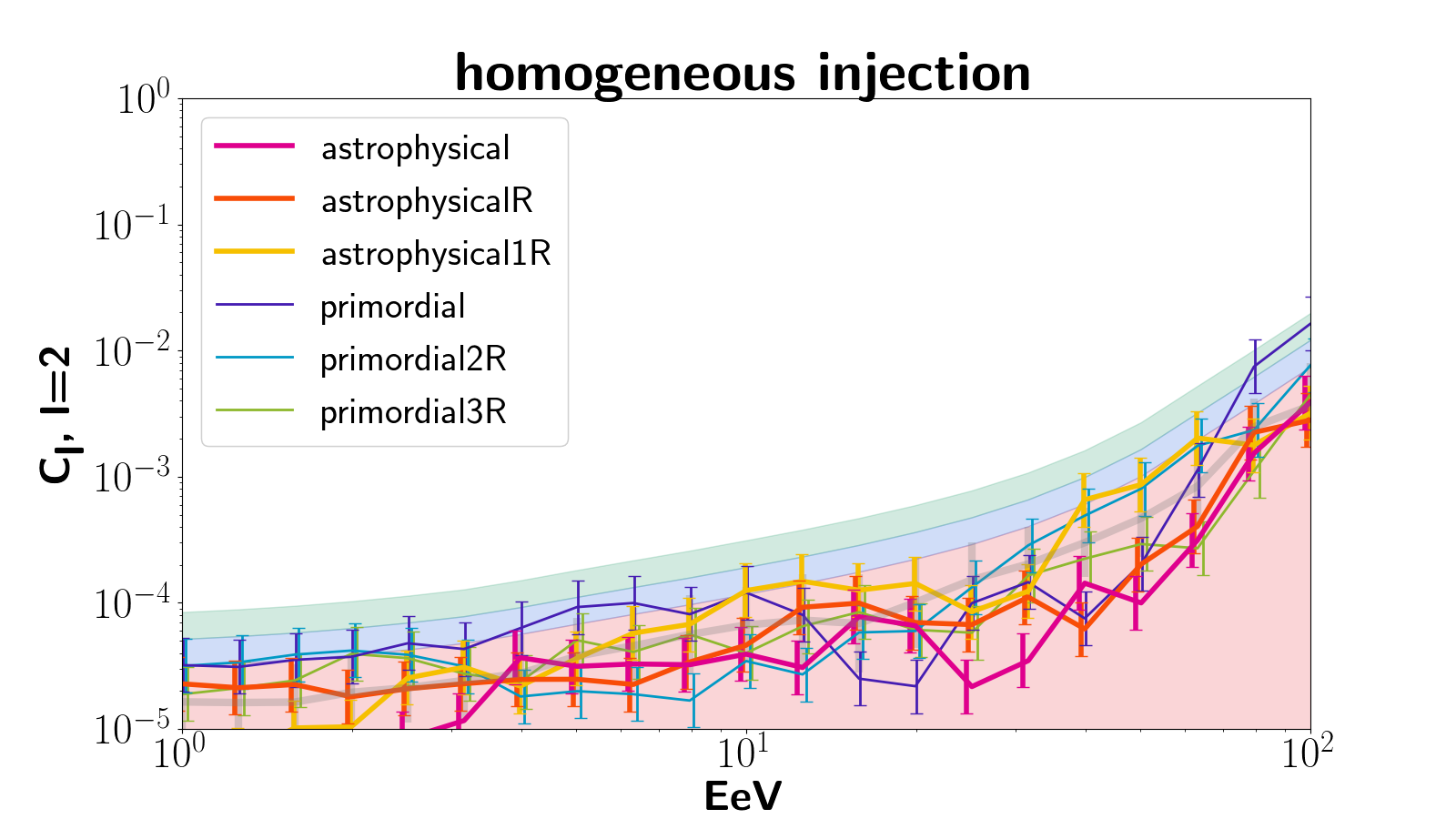}
\includegraphics[width=0.5\textwidth]{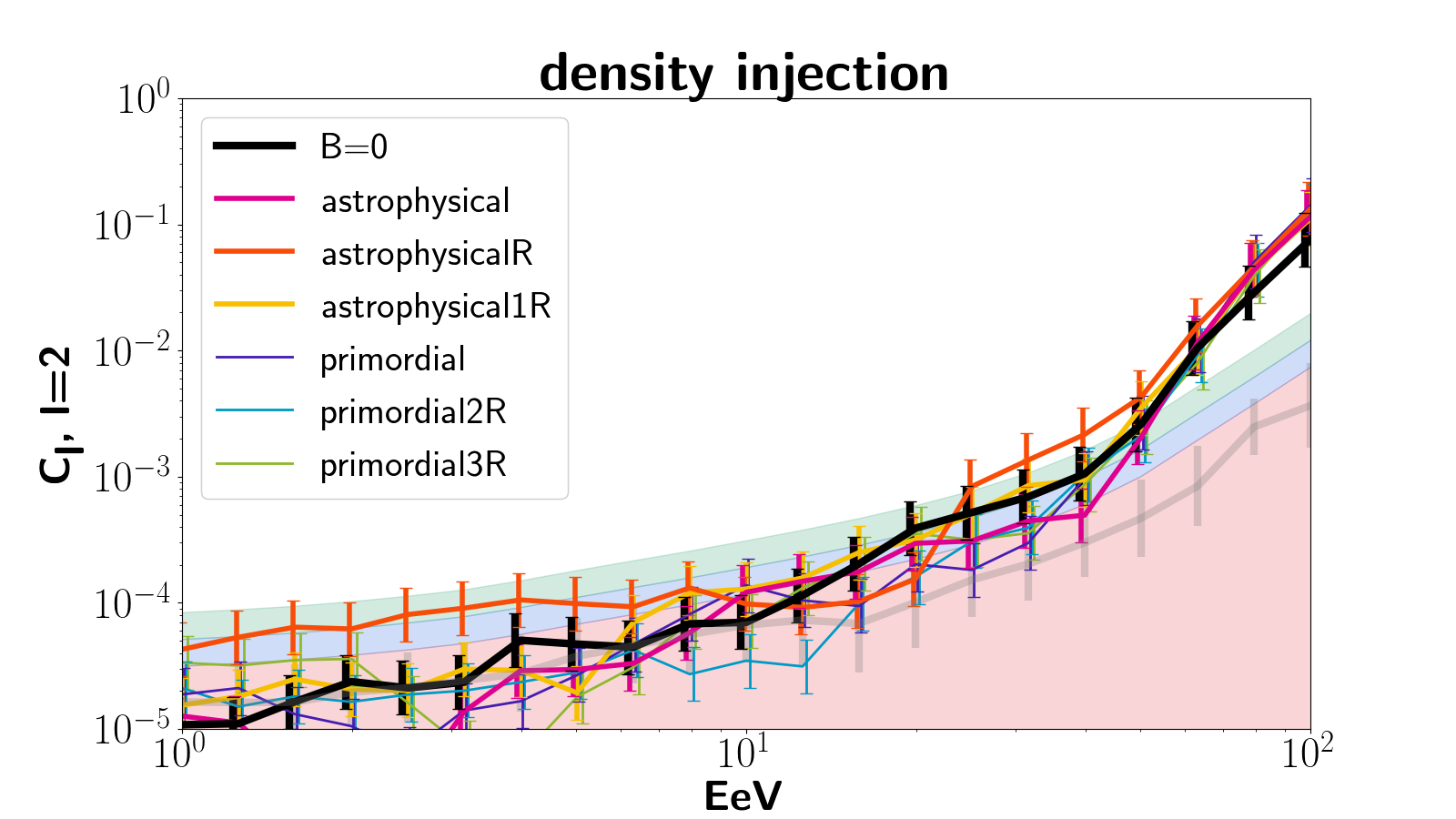}
\includegraphics[width=0.5\textwidth]{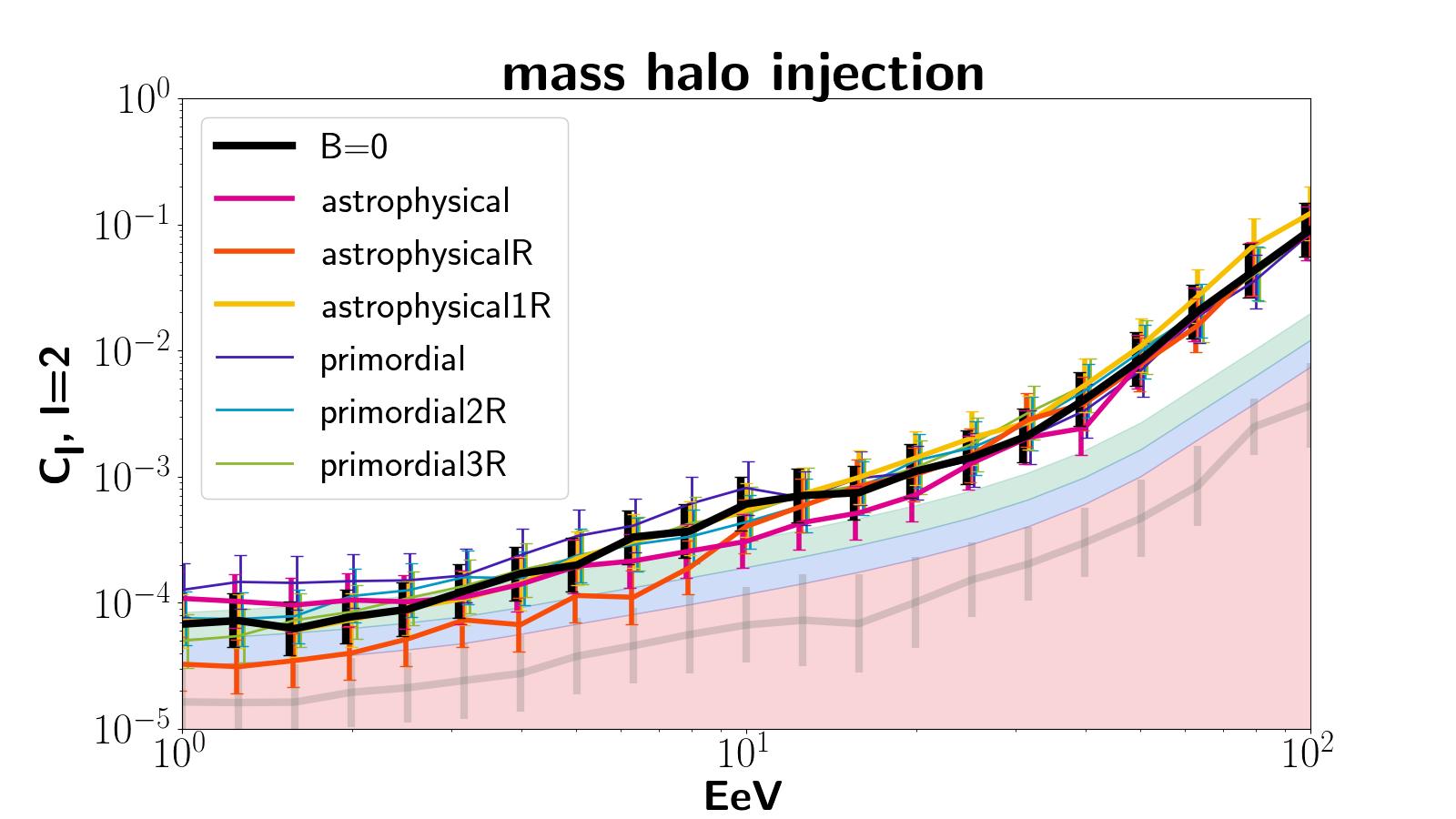}
\caption{ Angular power $C_l$ of the quadrupole $l=2$ for all models listed in Tab. \ref{tab:model} in a pure proton injection scenario.
\bff{	The error-bars indicate sample deviation given by Eq. \ref{eq:APS_variation}. }
	From top to bottom, the panels show the cases of \isotropic, \LSS and \mass injection listed in Tab. \ref{tab:source}.
	The thick grey line is the average and 1$\sigma$ standard deviation of the baseline homogeneous model.
	The shaded regions indicate the 68\%, 95\% and 99\% C. L. of anisotropy.}
\label{fig:APS}
\end{figure}
\begin{figure}
\includegraphics[width=0.5\textwidth]{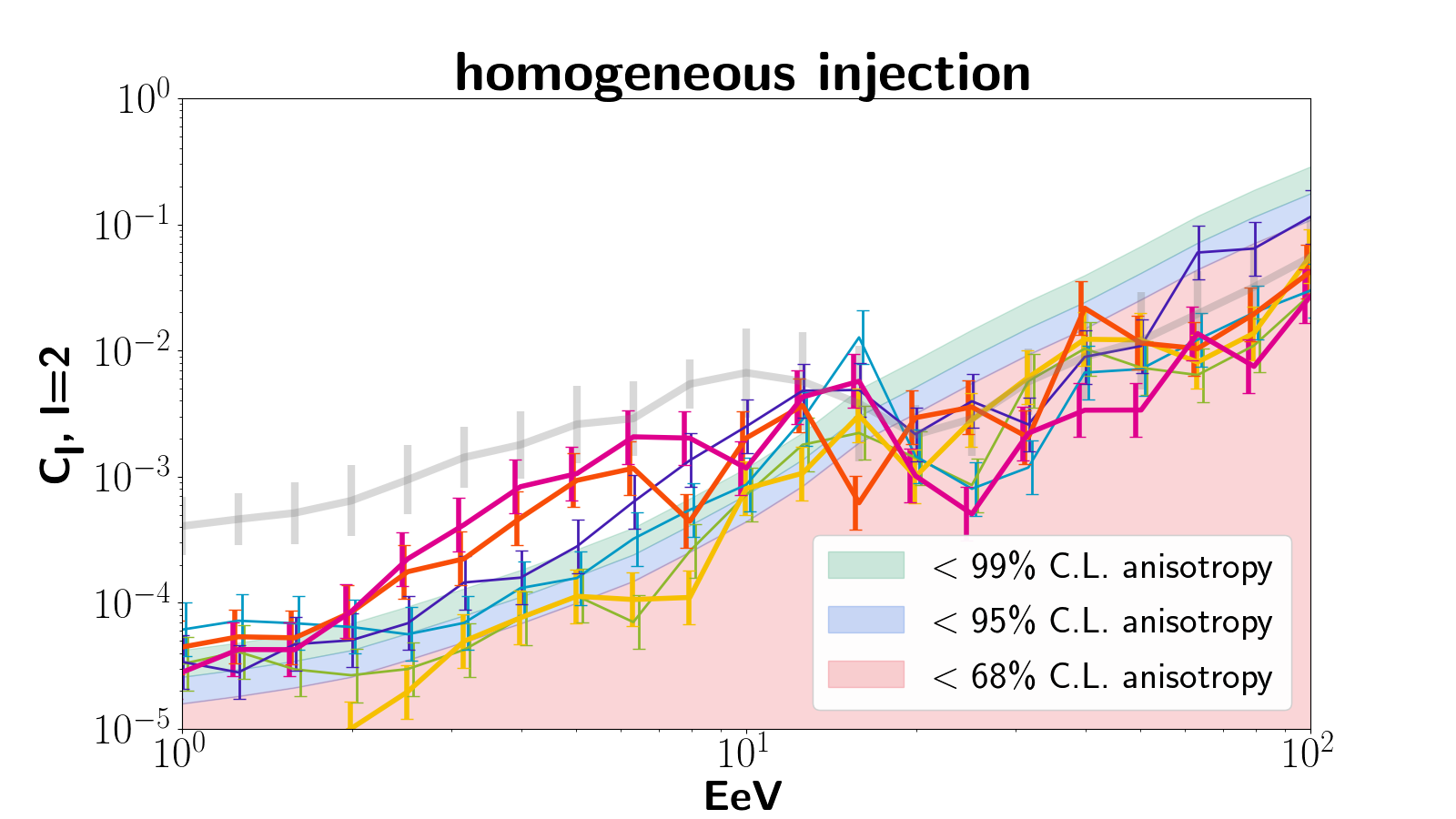}
\includegraphics[width=0.5\textwidth]{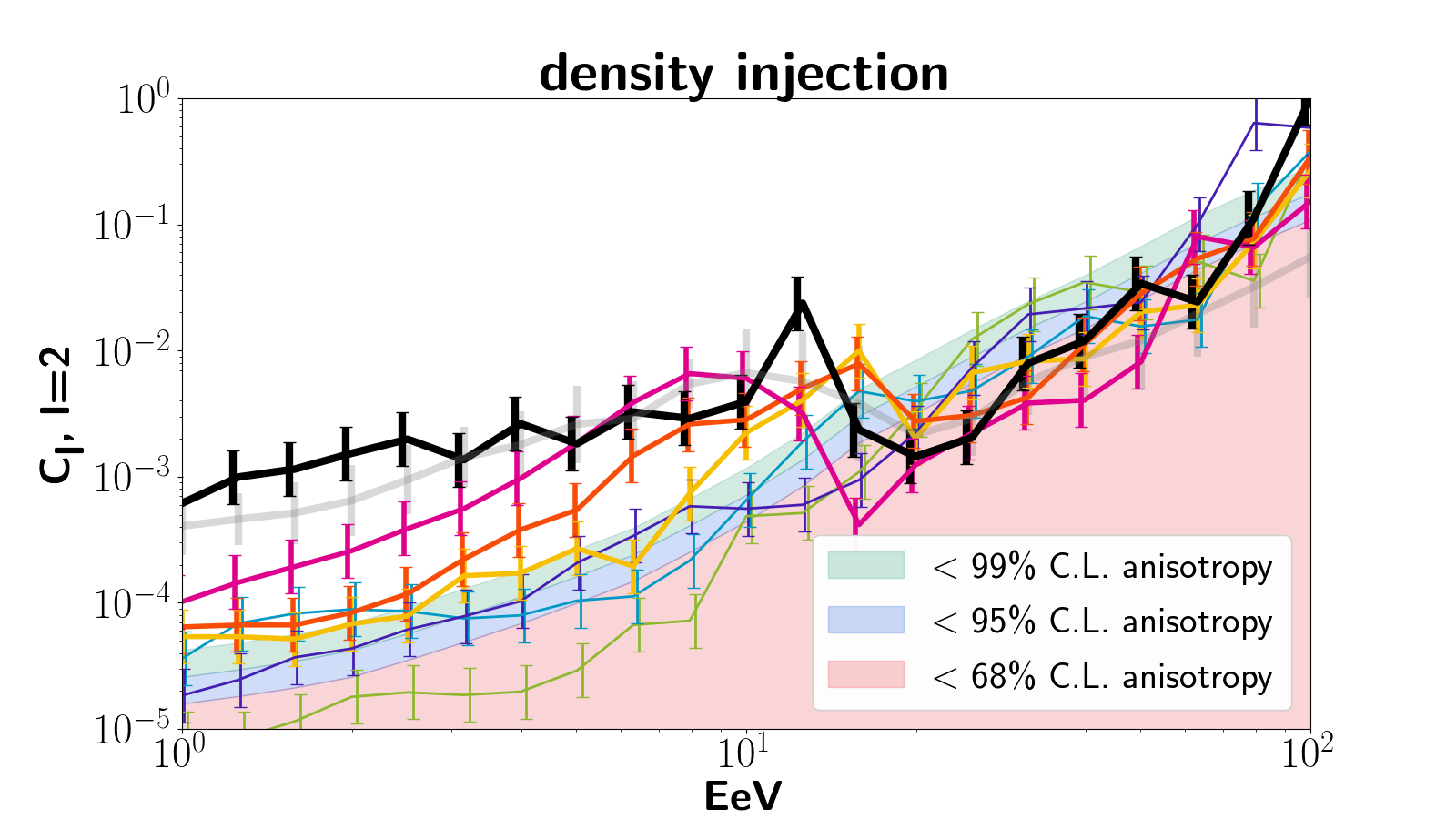}
\includegraphics[width=0.5\textwidth]{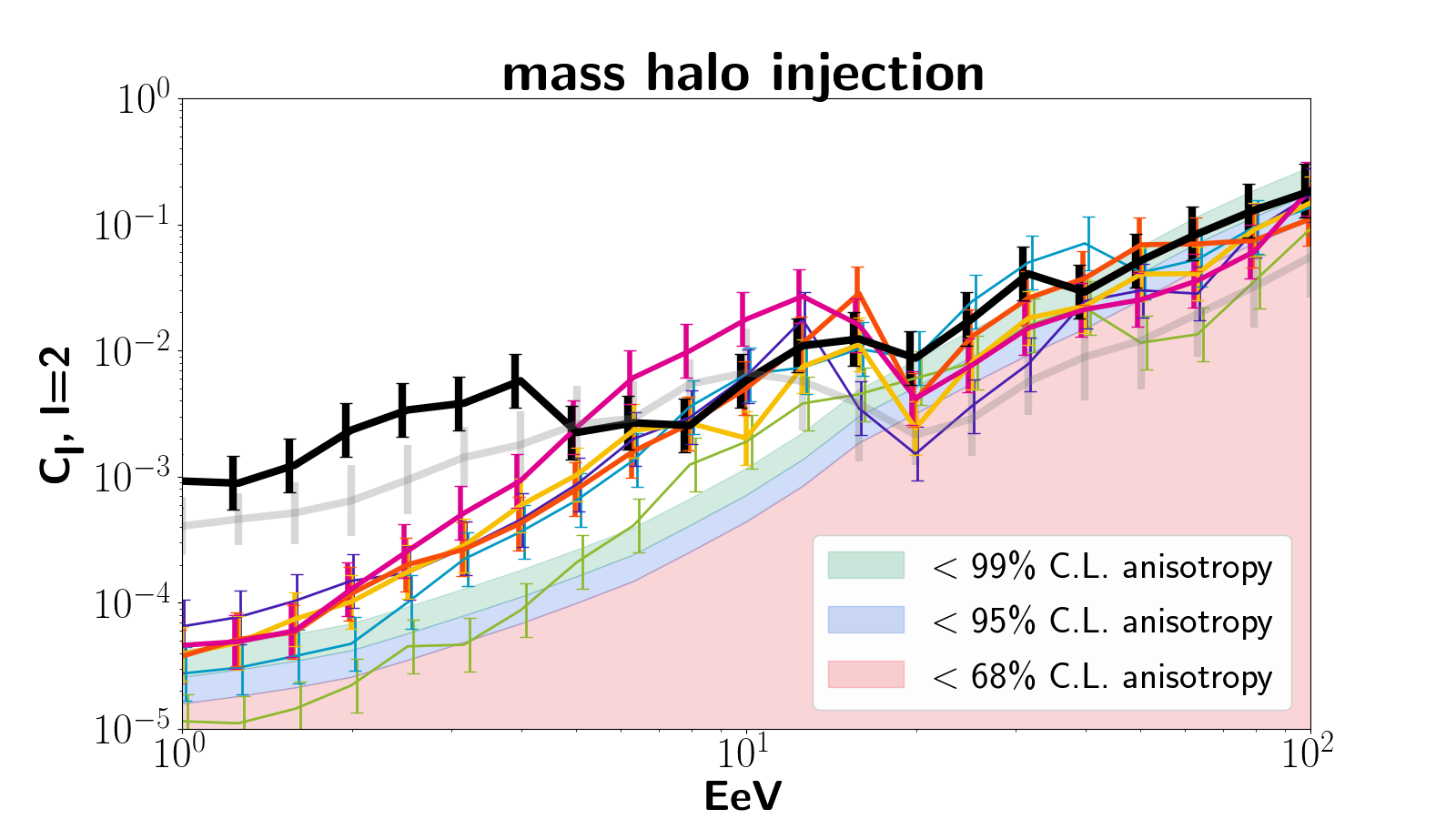}
\caption{ Same as Fig. \ref{fig:APS}, in a pure iron injection scenario.}
\label{fig:APS_iron}
\end{figure}
\begin{figure}
\includegraphics[width=0.5\textwidth]{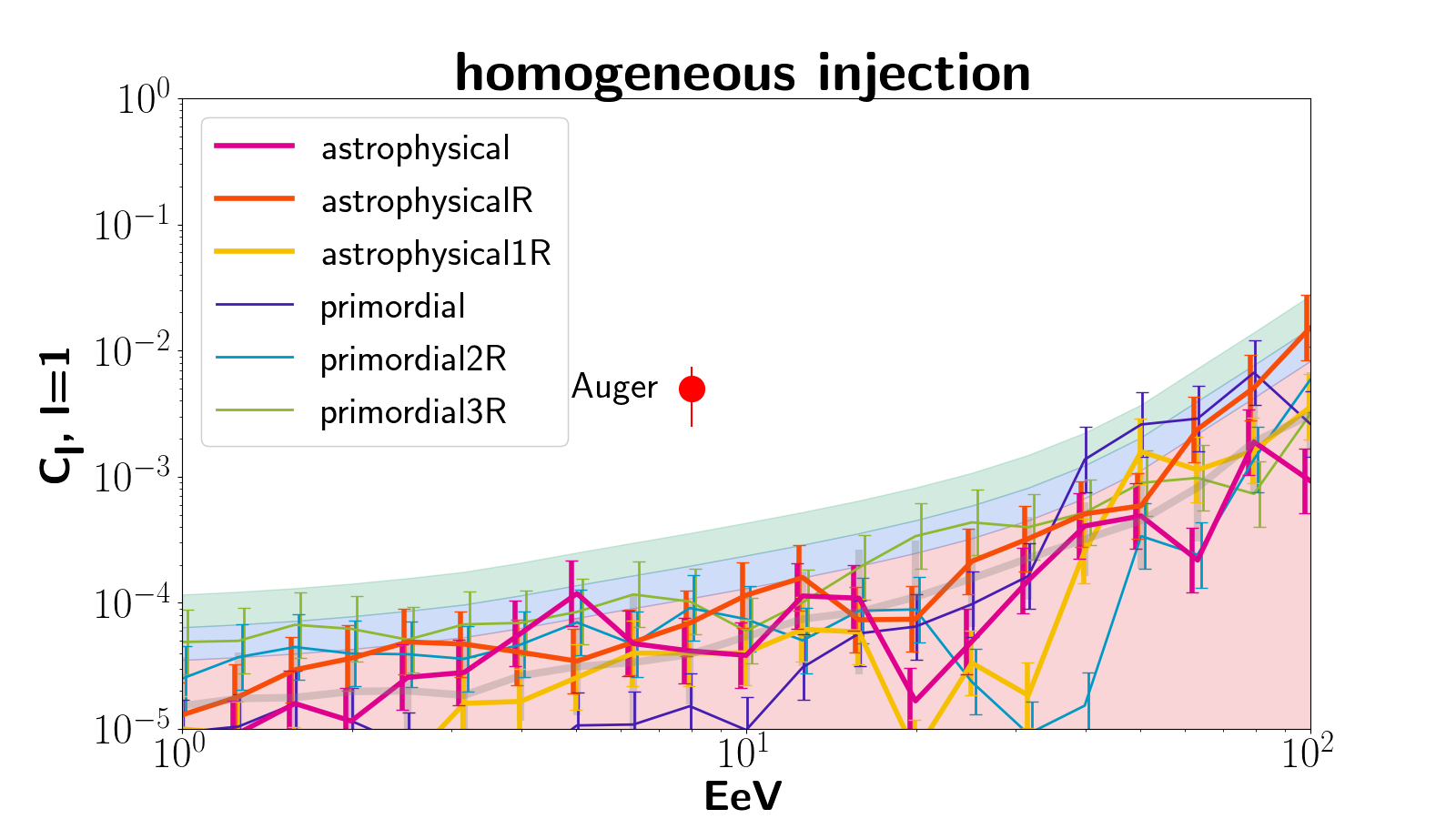}
\includegraphics[width=0.5\textwidth]{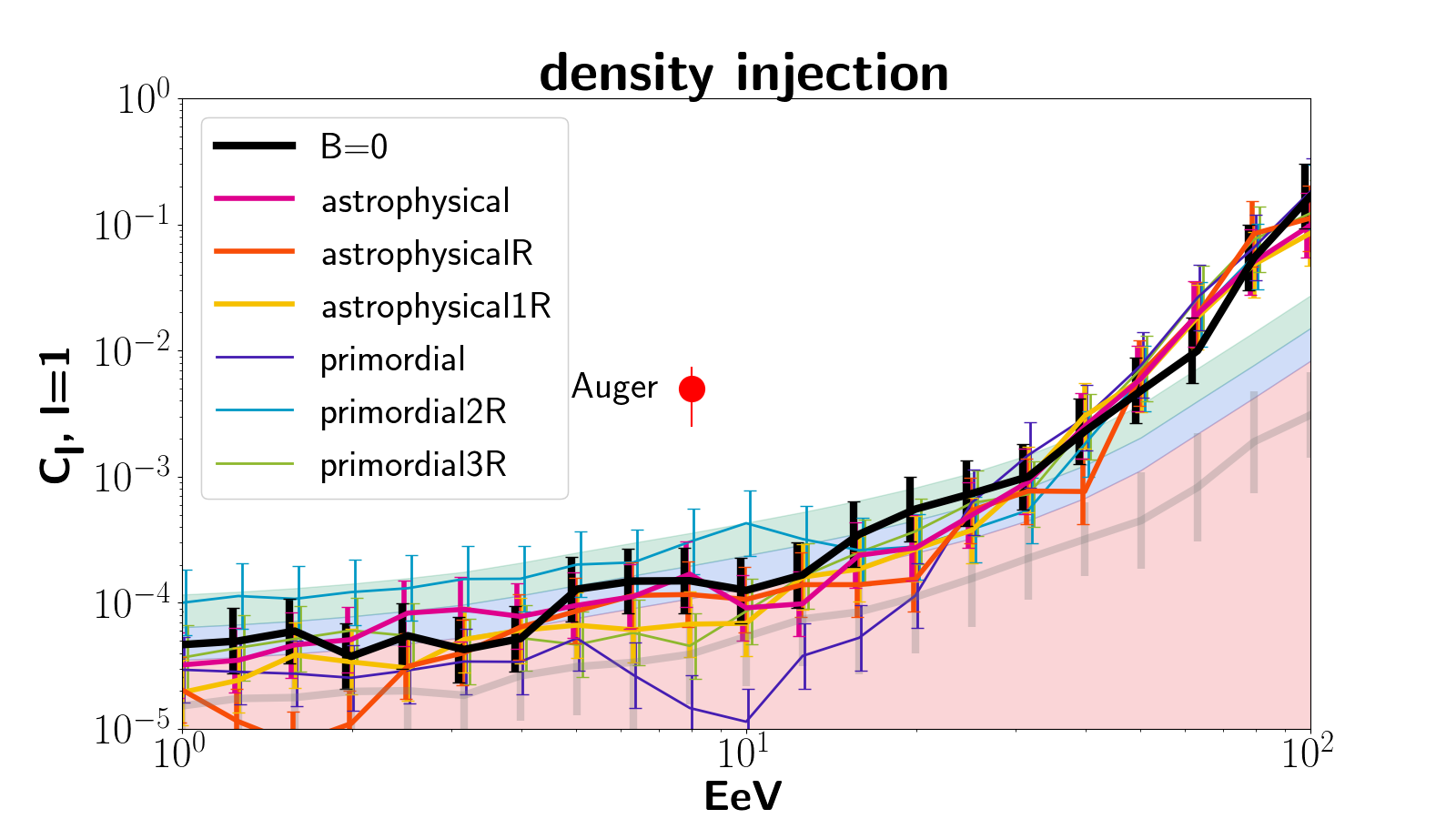}
\includegraphics[width=0.5\textwidth]{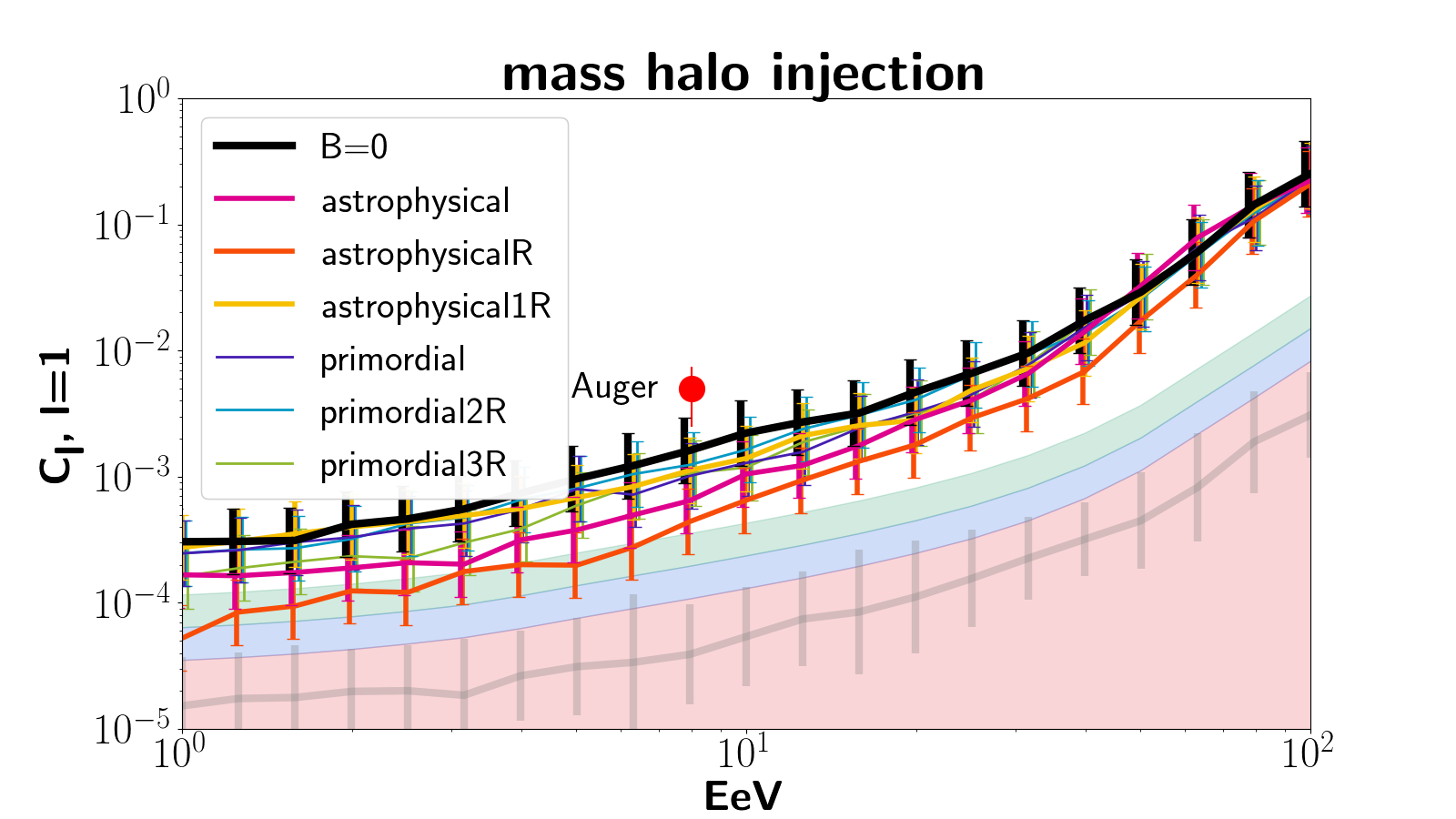}
\caption{ Angular power $C_l$ of the dipole $l=1$ for all models listed in Tab. \ref{tab:model} in a pure proton injection scenario.
	The error-bars indicate sample deviation given by Eq. \ref{eq:APS_variation}.
	From top to bottom, the panels show the cases of \isotropic, \LSS and \mass injection listed in Tab. \ref{tab:source}.
	The thick grey line is the average and 1$\sigma$ standard deviation of the baseline homogeneous model.
	The shaded regions indicate the 68\%, 95\% and 99\% C. L. of anisotropy.
	The red point corresponds to the amplitude of the recent dipole signal reported by Auger.}
\label{fig:APS_dip}
\end{figure}
\begin{figure}
\includegraphics[width=0.5\textwidth]{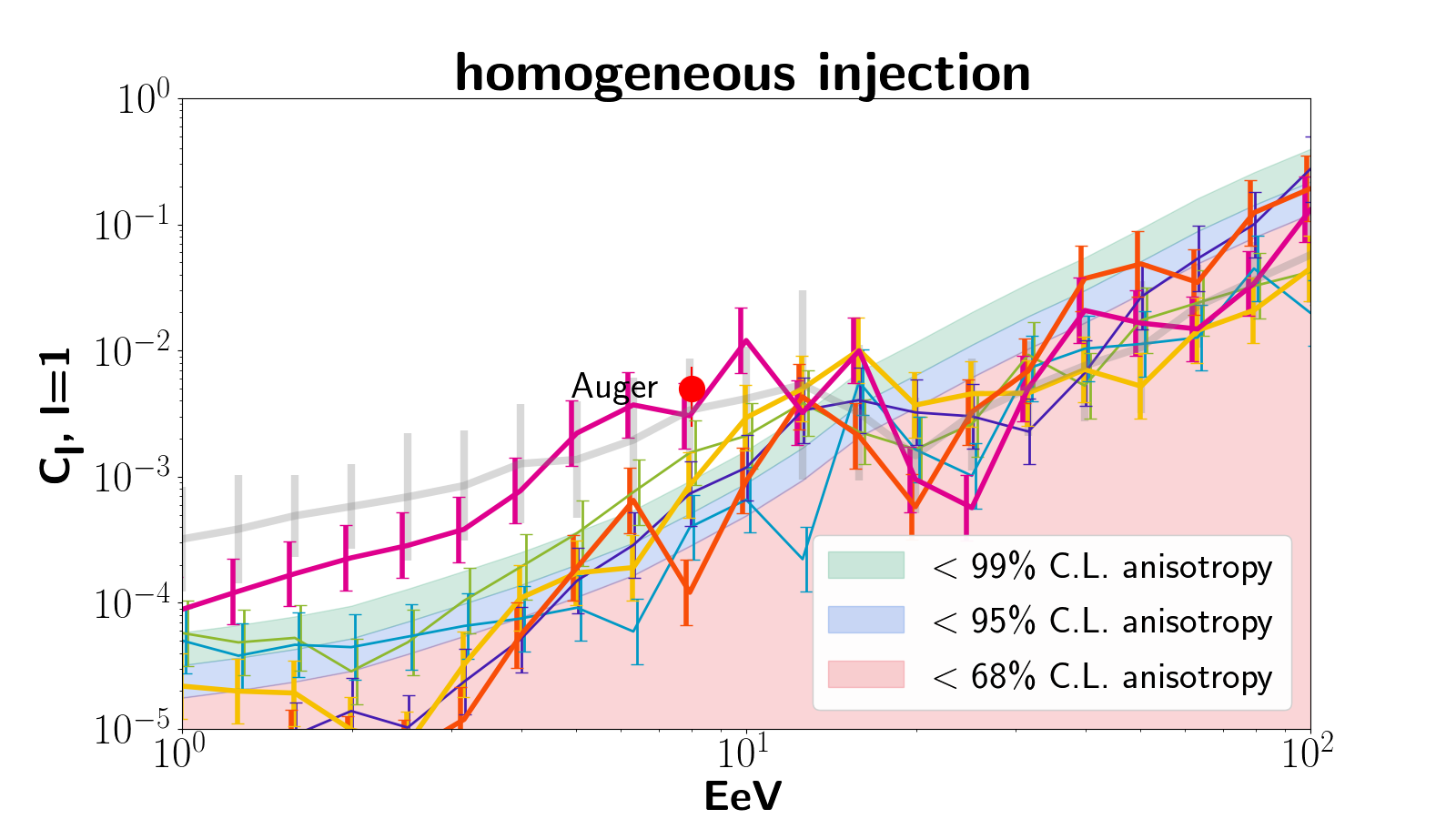}
\includegraphics[width=0.5\textwidth]{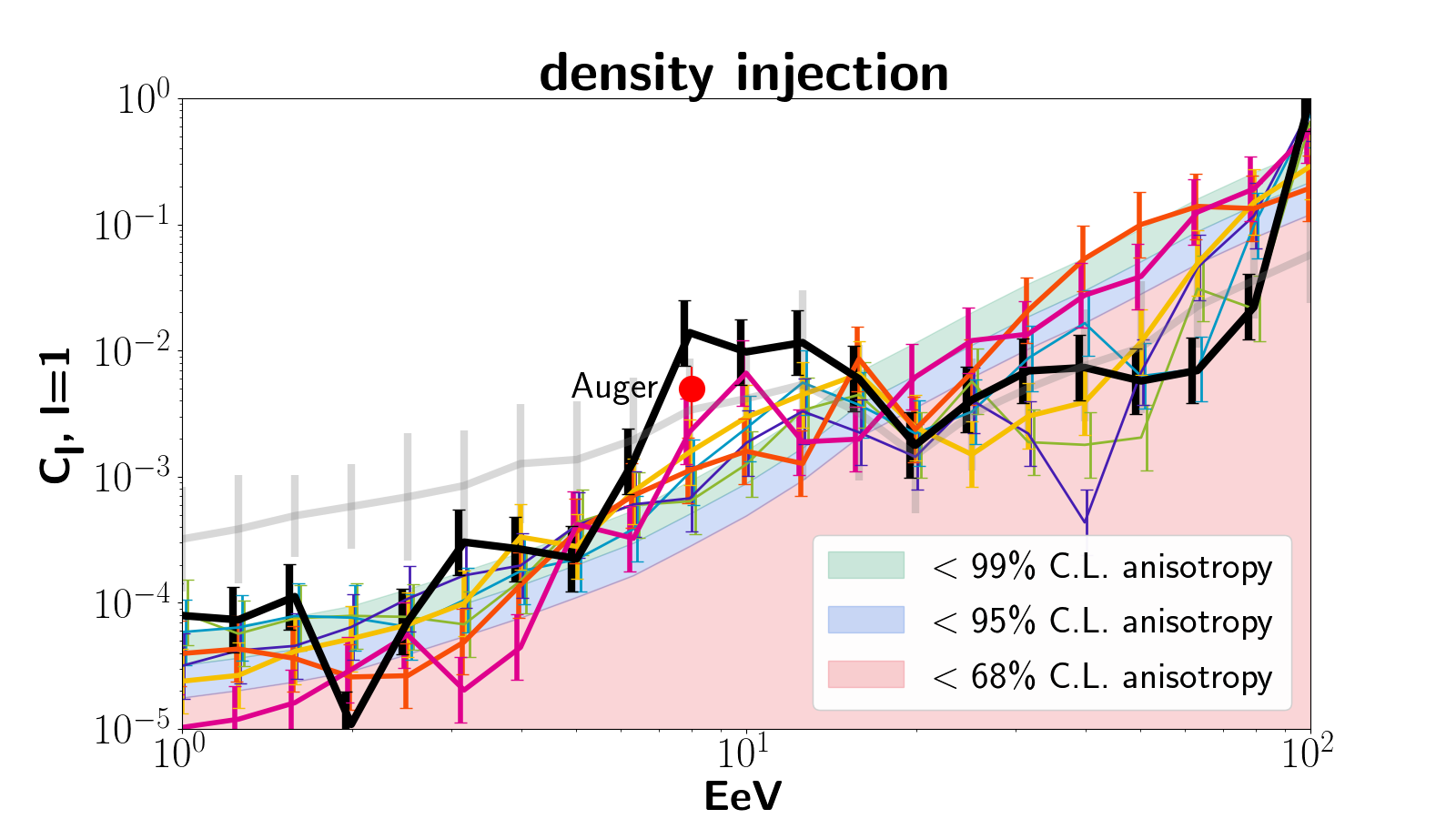}
\includegraphics[width=0.5\textwidth]{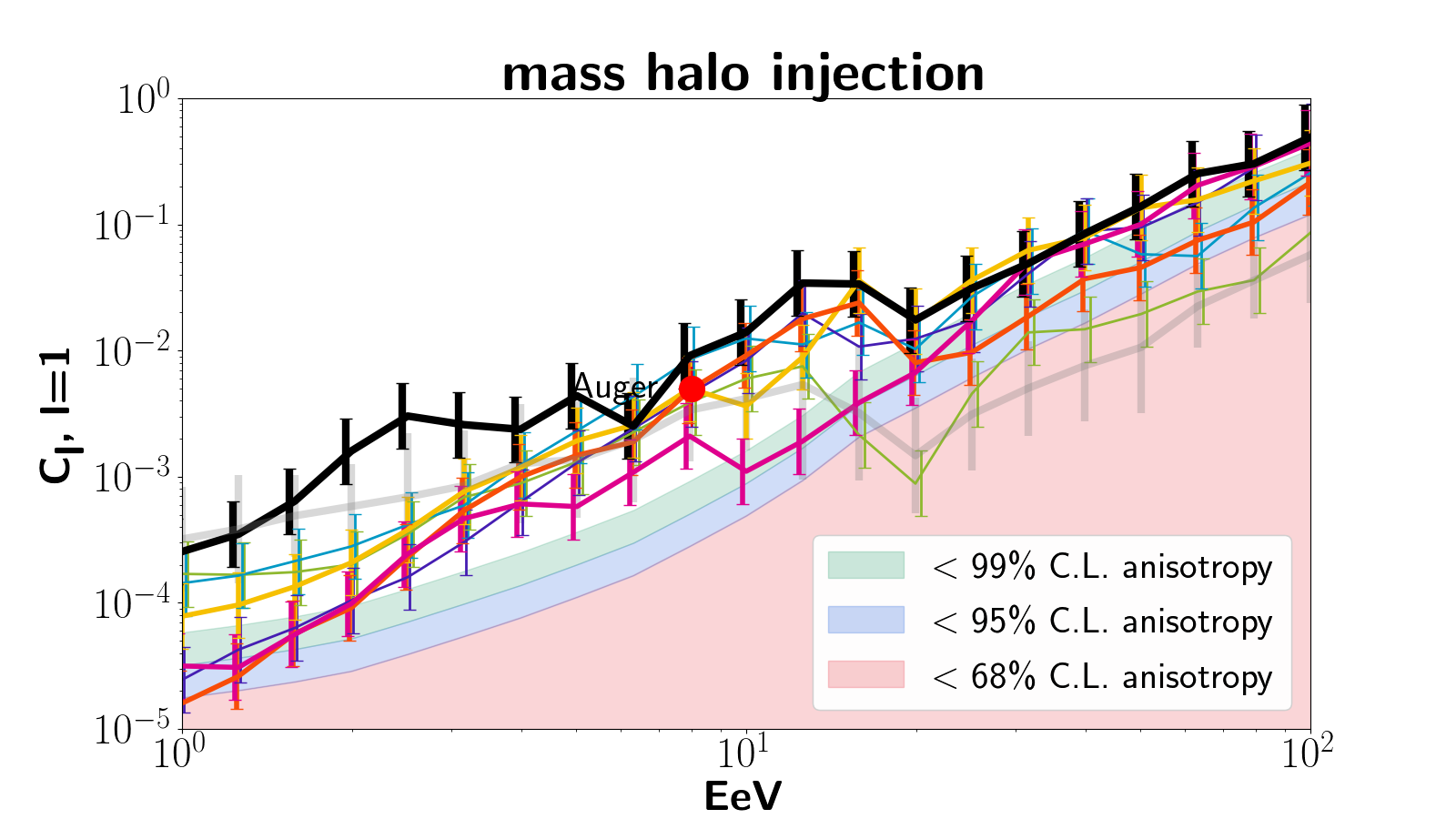}
\caption{ Same as Fig. \ref{fig:APS_dip}, in a pure iron injection scenario.}
\label{fig:APS_dip_iron}
\end{figure}
	To compute the angular power $C_l$ presented in this section, we first produce full-sky maps of the arrival directions of UHECR events for different minimum energies of considered particles.\footnote{\rbff{Due to the hard injection spectrum used in our simulations the full-sky maps contain too many events at high energies. However, since the observed spectra in general are steeper than $E^{-1}$, this effect is negligible. }}
	These maps are then decomposed into spherical harmonics $ \Phi(n) = \sum a_{lm} Y_{lm}(n)$ and $C_l$ is calculated from the obtained amplitudes, $C_l = (2l+1)^{-1} \sum |a_{lm}|^2$ \citep[cf. e. g.][]{2015JETP..120..533T}.
\bff{ 	Finally the whole spectrum is normalized by the monopole moment, which is 4$\pi$ times the square of the average flux.}
\\ \\
	We present the dipole and quadrupole moment of the angular power spectrum $C_l$ of UHECR arrival directions.
	These moments were shown to be most promising in the search for anisotropy signals \citep{2017arXiv170602534D},
	but the general trends reported in this section also apply to the octopole moment.
\\ \\
	The isotropic prediction is obtained analytically for an isotropic full-sky with $N$ events \citep{2015MNRAS.448.2854C} . 
	The mean value of the angular power
\begin{equation}
		C_l = 4 \pi / N
	\label{eq:APS_isotropic}
\end{equation}
	and the general sample deviation
\begin{equation}
	\sigma = \sqrt{ \frac{2}{2l + 1} } C_l \  ,
	\label{eq:APS_variation}
\end{equation}
\bbff{which shows the variation for realizations of a Gaussian random process.}
	For an isotropic sky, both, $C_l$ and $\sigma$, scale with $1/N$.
	The logarithmic deviation stays constant.
	In order to account for fluctuation in $C_l$, we show $\sigma$ as error bars for every graph.
\\
	Since the value and fluctuation of $C_l$ in an isotropic sky of finite counts are determined by the number of events and
	our simulations do not reproduce the spectrum observed in nature (cf. Sec. \ref{sec:energy_spectrum}), we need to compare to predictions for the simulated spectrum which depends on the injected composition.
\bbff{We indicate with shaded regions the confidence level of anisotropy (C. L. anisotropy).
	This is obtained from the isotropic prediction and 1, 2 \& 3$\sigma$ sample deviation, equations (\ref{eq:APS_isotropic}) and (\ref{eq:APS_variation}).
	The number of particles $N$ used to calculate the isotropic prediction is the average $N$ observed in each energy bin.
	The fluctuation of $N$ is about 10\% for same injection composition, so the C. L. anisotropy is roughly the same for all models.
}
\\
	In addition to the simulations with magnetic field in all plots we also present a simulation where the magnetic field is globally set to zero, \bzero\ (black line).\bbff{\footnote{except for the \isotropic plot, where it is given by the homogeneous prediction.}}
	This simulation is shown in order to unambiguously determine the cases where the magnetic field model is important.
\\ \\
\bbff{We further show the prediction given by a baseline homogeneous model (thick grey line).}
	It shows the average and 1$\sigma$ standard deviation of a test group of 27 realizations of a scenario with \isotropic injection in the absence of magnetic fields, \bzero.
	These fully homogeneous scenarios produce the most isotropic results possible in our simulation.
	The result is not fully isotropic, since it entails all artifacts intrinsic in the simulation, e. g. finite observer effect, over-count of secondary nuclei and assumed periodicity of the magnetic field and sources \citep[for a detailed discussion see][]{finite_observer,Hackstein2016}.
\bbff{This makes the homogeneous baseline model a suitable test to find the qualitative contribution of sources and magnetic fields.}
\\ \\
	The proton injection scenarios are shown in Fig. \ref{fig:APS}.
\bbff{The prediction from the homogeneous baseline model obtained by the procedure explained above is almost identical to the isotropic prediction.}
	At energies below the GZK-limit of $\sim 40 \rm\ EeV$, 
	the quadrupole angular power is in good agreement with the isotropic prediction for the \isotropic and \LSS injection models presented in the top two panels.
	In the \mass injection model the angular power is above 95\% C. L. anisotropy at all energies in virtually all of the models.
	This is in agreement with results from \citet{2017arXiv170602534D} and \citet{Abreu:2013kif}
	that show that UHECRs cannot predominantly be protons from few sources in the LSS
	and that an anisotropic signal should have already been measured for source densities $\lesssim 10^{-4} \rm\ Mpc^{-3}$.
\\
	The magnetic field models do not significantly change the angular power spectrum of arrival directions of UHECR protons at all energies.
	At very high  energies, $\sim 100 \rm\ EeV$, the variation in the coefficients, $C_l$, between the magnetic field models is the lowest,
	though the number of protons and thus the accuracy is the lowest.
	The \LSS and \mass injection models show a strong deviation from isotropy, while the \isotropic injection is in good agreement with the prediction from isotropy.
\bff{The error bars indicate that this feature is not an effect of sample variance, but is statistically significant.}
	This shows that the distribution of nearby sources imposes on the observer an anisotropic signal of UHECRs right below the energy cut-off, where propagation of UHECRs is believed to be quasi-rectilinear.
	This anisotropic signal can be used to identify the sources of UHECRs.
\\ \\
\bbff{	In the iron injection scenario shown in Fig. \ref{fig:APS_iron}, almost all models have significantly higher values of $C_l$ below $20 \rm\ EeV$ than expected in an isotropic distribution.
	This energy coincides with $E_{\rm max} / A_{\rm Fe}$, the maximum energy of injected particles $E_{\rm max} = 1000$ EeV divided by the mass number of iron $A_{\rm Fe} = 56$.
	The prediction from the homogeneous baseline model and the \bzero\ model generally show the highest values.
	Anisotropy occurs, independent of the source model, due to complete disintegration of heavy nuclei over very short length scales after they have been injected nearby at the highest energies.
	Due to the high Lorentz-factor, in the absence of deflection the arrival directions of these secondary nuclei are almost identical,
	causing an excess of events in direction of the most nearby injection positions \citep[cf. e. g. ][]{2009JCAP...11..009L}.
	We see that the stronger \prim models in general show lower $C_l$ values than the weaker \agn models.
	The anisotropy produced by the procedure explained above is lowered by CMFs.
	Since the anisotropy is predominantly produced by nearby sources, only the local field \citep[up to $10\times$ distance to closest source,][]{Dundovic:2017vsz} is responsible for this effect.
	This is in agreement with \citet{2004PhRvD..70d3007S}, who infer that strong magnetic fields around the observer can suppress large-scale anisotropy.
\\
	At the highest energies, $\sim$ 100 EeV, the number of observed events is too low in the iron injection case to measure the deviation from isotropy.
}
\\ \\
\bbff{	While this manuscript was under review, the Pierre Auger Collaboration reported a significant dipole in the arrival directions of UHECRs with energies $> 8$ EeV at a $5.2\sigma$ level of significance with an amplitude of $6.5 \%$ \citep{2017arXiv170907321T} or $C_1 = 0.0050 \pm 0.0025$ in terms of the angular dipole power \citep{2017JCAP...06..026A}.}
	In Figs. \ref{fig:APS_dip} and \ref{fig:APS_dip_iron} we present the dipole moment $l=1$ of the angular power $C_l$ in our simulations and also indicate the recent observation.
	The features in these graphs are basically the same as discussed for the quadrupole.
\\
	Note that the number of particles above 8 EeV in our simulations is different from the amount of particles considered in \citet{2017arXiv170907321T}.
	While their result is calculated for $\gtrsim 32.000$ events $> 8$ EeV, our simulations have only about $\sim 17.000$ and $\sim 6.000$ events in the proton and iron runs, respectively.
\bbff{Hence results of our simulations are of lower statistical significance.}
	However, the energy spectra in our simulations are much harder than observed by the Pierre Auger Collaboration and therefore anisotropic signal from source distribution are expected to be more dominant.
\\
\bbff{	None of the models explored in this paper can reproduce the signal observed  in nature with pure proton injection (Fig. \ref{fig:APS_dip}).
	Only for the \mass model there is a small overlap of 1$\sigma$ deviations with the Auger measurement.
	The level of anisotropy does not decrease strongly from 8 EeV to 4 EeV.
	This indicates that a strong dipole in the distribution of nearby sources is necessary to reproduce the Auger signal with a light injection composition.
	In such a scenario, a significant dipole is expected also in the 4 - 8 EeV energy bin, which is not observed in nature.
	This makes a light injection composition of UHECRs at the highest energies unlikely in view of the recent observation.
\\ \\	
	Injection of iron nuclei, Fig. \ref{fig:APS_dip_iron}, results in a dipole similar to that observed by Auger - in amplitude, not in significance.
	In the \isotropic model magnetic fields can suppress the signal to agree with the isotropic prediction.
	In the \LSS and \mass model, the anisotropic signal is dominated by the distribution of sources and not suppressed efficiently by magnetic fields.
\\
	Our results suggest that the dipole signal in UHECRs observed by the Pierre Auger Observatory may be the product of clustering of secondary nuclei in direction of the nearby sources.
}
\subsection{Composition}
\label{sec:composition}
\begin{figure}
\includegraphics[width=0.5\textwidth]{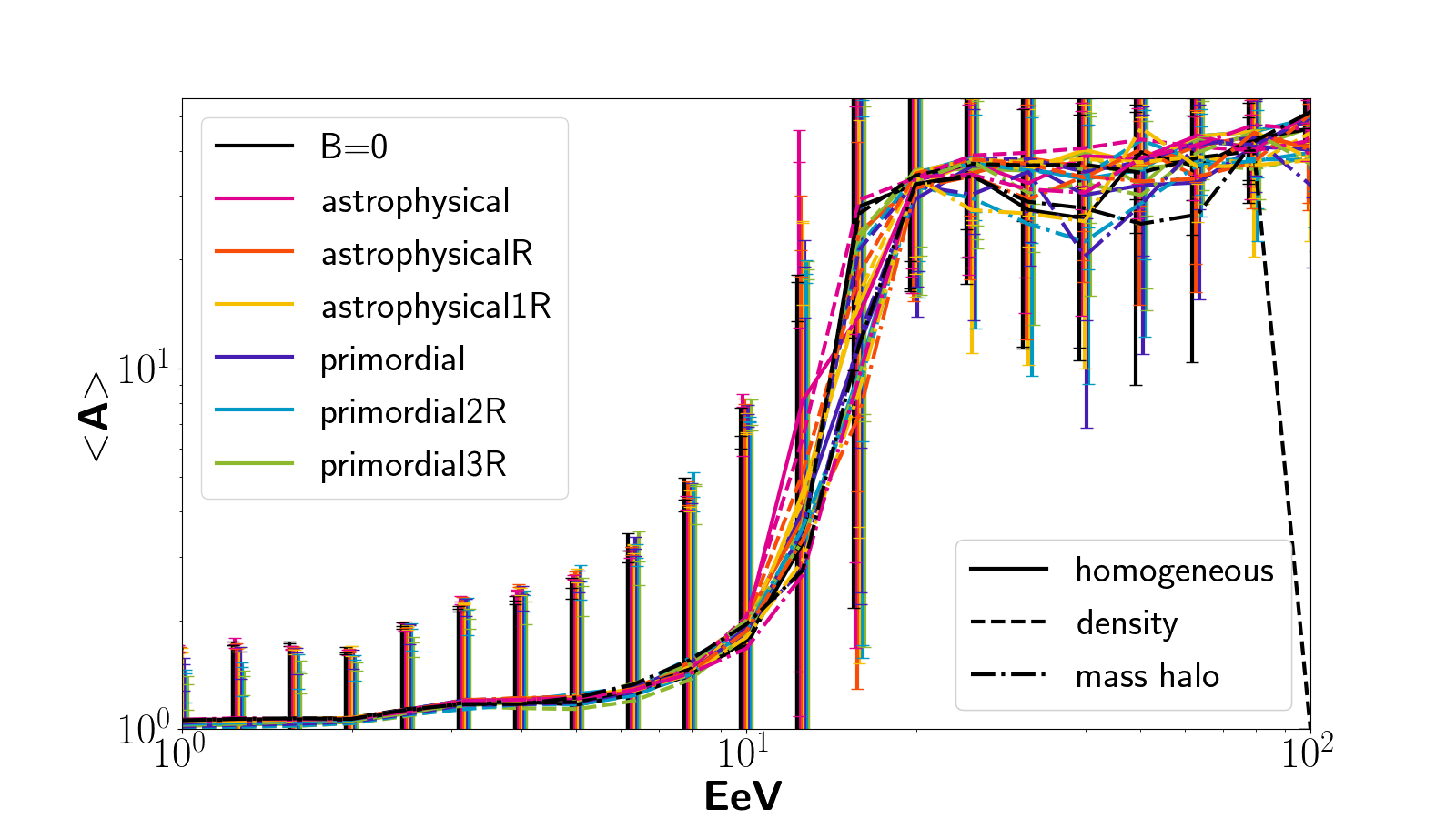}
\caption{ Average mass number $\langle A \rangle$ of UHECRs observed at different energies.
	The errorbars show the 1$\sigma$ standard deviation.
	The colours indicate the magnetic field model listed in Tab. \ref{tab:model} and the linestyle shows the source model listed in Tab. \ref{tab:source}.
	}
\label{fig:composition}
\end{figure}
	In Fig. \ref{fig:composition}, we show the average mass number $\langle A \rangle$ of observed events as function of energy in the iron injection scenarios.
	At low energies, the composition is very light since secondary protons of injected iron nuclei dominate observations.
	At $\lesssim 20$ EeV $\approx E_{\rm max} / Z_{\rm Fe}$ there is a steep increase in $\langle A \rangle$.
	This coincides with the maximum energy of secondary protons.
	At higher energies only the (partly disintegrated) primary nuclei are observed.
	All magnetic field and injection models show a very similar slope of $\langle A \rangle$.
	We conclude that CMFs in agreement with observational upper limits in general are too weak to impose a significant difference in the all-sky average composition of UHECRs.

%% file: conclusions.tex
\label{sec:conclusion}
	We have studied the influence of CMFs on the propagation of UHECRs using MHD-simulations with different models for seeding of magnetic fields for both, primordial and astrophysical processes.
	We have found no evidence that magnetic field seeding scenarios could be distinguished via the use of the angular power spectrum of the spherical harmonics decomposition of the full-sky of arrival directions of UHECRs. 
\\ \\
	We have studied the influence of different source scenarios on the energy spectrum of UHECRs and on the angular power of  anisotropy.
	We have found that for a pure proton composition the slope of the energy spectrum at energies $> 100$ EeV depends on the number of, and distance to, the most nearby sources.
	The closer the sources, the harder the energy spectrum.
	If only iron is injected, almost no events are observed above that energy.
	Thus, the sharp energy cut-off observed with extensive air shower arrays \citep{2010ApJ...712..746I,2014BrJPh..44..560L} might suggest a low number of sources in the near vicinity of the observer if the cut-off does not coincide with the maximum energy of proton acceleration.
\\ \\
	We have investigated the angular power spectrum of arrival directions. 
	We have found  that there is a clear deviation from isotropy, $\lesssim 100$ EeV, if the distribution of sources follows the LSS.
	This offers the chance to identify the sources with future full-sky measurements \citep{2017arXiv170307897D} and high number statistics at the highest energies.
\\
\bbff{	We were able to reproduce the dipole in the arrival directions of UHECRs $>8$ EeV recently reported by the Pierre Auger Collaboration \citep{2017arXiv170907321T} with all our source models, but only using pure iron injection composition instead of protons.
	Our results indicate that the observed dipole is the result of clustering in direction of nearby sources of heavy nuclei \citep[][]{2009JCAP...11..009L}.
	Strong magnetic fields might be necessary to explain the absence of a dipole in the 4-8 EeV energy bin.
	Exploring such possibilities (also joined with a more thorough exploration of the role of UHECRs composition in the production of a dipole excess) will be subject of forthcoming work.
}
\\ \\
	For the injection of protons from the virial halos with a very low number density, around the limit from \citet[][$\sim 10^{-4} \rm Mpc^{-3}$]{Abreu:2013kif},
	we have found 95\% C. L. quadrupolar anisotropy at all energies, in conflict with present observations \citep{2012ApJS..203...34P,2014ApJ...794..172A}.
	This confirms the findings of \citet{2017arXiv170602534D} that UHECRs cannot primarily be protons from few sources in the LSS.
\\ 
	Finally, we have analyzed the observed composition of UHECRs via the average mass number of events.
	There is no evidence that CMFs significantly influence the all-sky composition of UHECRs at all energies.
\\ \\
	In our study we did not account for the influence of the magnetic field of the Milky Way,
	but energy losses are negligible on galactic scales.
	Furthermore, the angular power spectrum at large scales has been shown to have low impact of deflections in the Galactic magnetic field \citep{2015JETP..120..533T,2017arXiv170602534D}.
\\ \\
	In summary, with newer constrained simulations of the local Universe we confirmed our previous findings \citep{Hackstein2016},
	i.e. that the properties of observed UHECRs do not seem to carry much information on the genesis and distribution of extragalactic magnetic fields.
	This in turn strengthens the possibility of performing "UHECRs astronomy"  \citep{Dolag:2003ra}, 
	thus motivating further investigations on the origin of UHECRs across a wide range of energies where the impact of the Galactic magnetic field should be sub-dominant. 

%% file: appendix.tex
\section{Re-weighted energy spectrum}
\bbff{
\begin{figure*}
\includegraphics[width=0.495\textwidth]{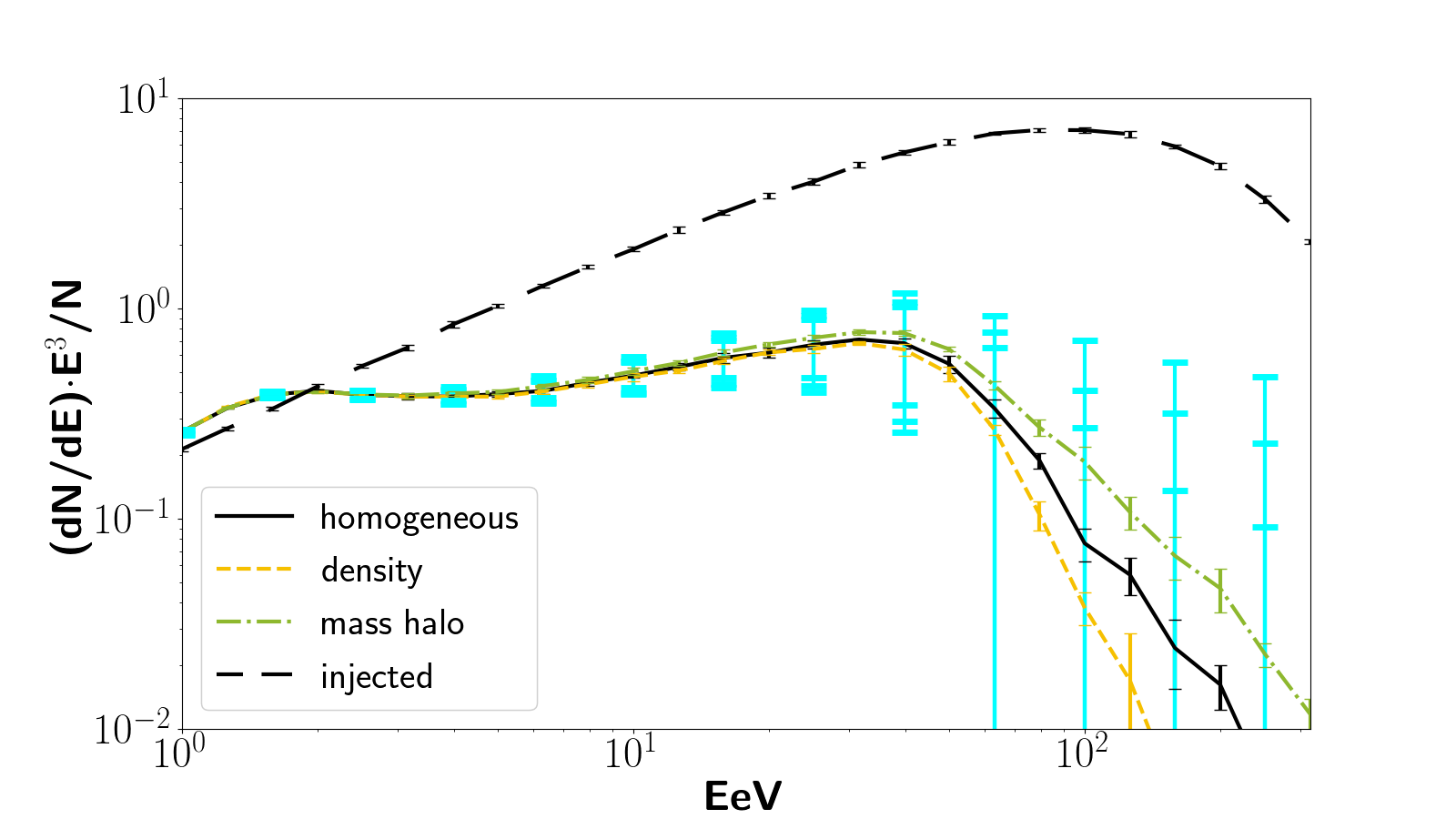}
\includegraphics[width=0.495\textwidth]{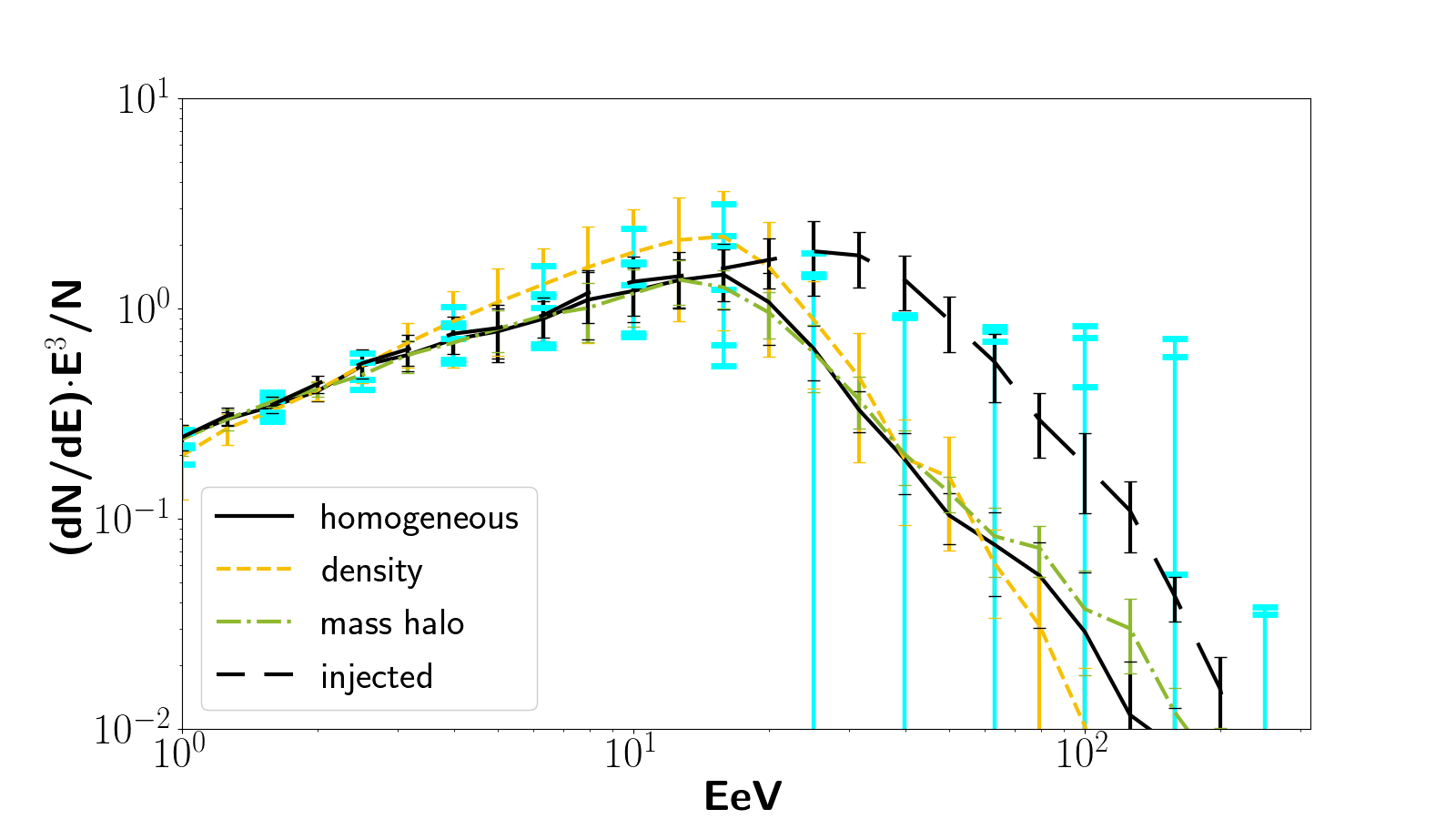}
\caption{Re-weighted energy spectrum of UHECRs as injected at the sources (dashed lines) and measured by the observer  for a pure proton and a pure iron injected composition (left and right, respectively).
	The colours and line styles indicate the injection models listed in Tab. \ref{tab:source}.	
	The graphs show the average over all magnetic field models, the standard deviation is indicated by the narrow error bars.
	The big cyan error bars show the Poisson noise at each second data point .
	The graphs are renormalized with the total number of particles $N$ and multiplied by  $E^{-3}$ to enable better comparison to the figures presented in \citet{2014BrJPh..44..560L}.
	}
\label{fig:spectrum_reweight}
\end{figure*}
	The energy spectra presented in this work do not recreate the spectrum observed in nature.
	This can be achieved by modifying the injection spectrum, in particular by using a softer spectral index and introducing an exponential cut-off.
	The resulting injection spectrum has the form
\begin{equation}
		\frac{\text{d}N}{\text{d}E} 
	\propto 
		E_0^{-\gamma} e^{-E_0 / ( Z_0 R_{\rm max} ) } \ ,
\end{equation}
	with initial energy $E_0$, initial charge number $Z_0$, spectral index $\gamma$,  and maximum rigidity $R_{\rm max} = E_{0,\rm max} / Z_{0}$.
	The modification can be done in post-processing by multiplying every event with a specific weight factor \citep{finite_observer,AvanVlietPhD}
\begin{equation}
		w(E_0, Z_0)
	=
		E_0^{\gamma_{\rm init}-\gamma} e^{-E_0 / ( Z_0 R_{\rm max} ) } \ ,
\end{equation}
	where $\gamma_{\rm init}$ is the spectral index used for the simulation.
\\ \\
	In order to obtain the correct spectrum at injection we fit the observed spectral index between the ankle $E_A \approx 5 \rm\ EeV$ and the cut-off $E_C \approx 20\rm\ EeV$, which is observed to be $\gamma = 2.63 \pm 0.04$ \citep{2014BrJPh..44..560L}.
	The best fit in the proton injection scenarios is an injection index of $\gamma = 2$, as expected for Fermi acceleration.
	The best fit for the iron injection scenarios is $\gamma = 2.4$.
\\
	The shape of the spectrum beyond the cut-off energy $E_C$ is recreated well by using $E_{0,\rm max} = 100 \rm\ EeV$ for the exponential cut-off.
	The maximum rigidity is then $R_{\rm max, p} = 100 \rm\ EV$ for the proton injection scenarios and  $R_{\rm max, Fe} = 100 / 26 \rm\ EV \approx 3.8 \rm\ EV$ for iron injection scenarios.
	The resulting spectra are shown in Fig. \ref{fig:spectrum_reweight}.
\\ \\
	After re-weight, the effective number of observed particles is $N_{\rm eff, p} \approx 7,000$ in the proton injection scenarios and $N_{\rm eff, Fe} \approx 600$ in the iron injection scenarios.
\rbff{
	The isotropic prediction for the angular power spectrum depends on the number of particles (see Eq. 1).
	Therefore, after re-weight the isotropic prediction increases everywhere by about an order of magnitude at least.
	Accordingly, the color bands are raised in Figs. 4-7.
	All re-weighted scenarios are below 68\% C. L. anisotropy at all energies.
}
	
}

%% file: paper.bbl
\begin{thebibliography}{87}
\expandafter\ifx\csname natexlab\endcsname\relax\def\natexlab#1{#1}\fi

\bibitem[{{Aab} {et~al}\mbox{.}(2014){Aab}, {Abreu}, {Aglietta}, {Ahn},
  {Samarai}, {Albuquerque}, {Allekotte}, {Allen}, {Allison}, {Almela}, \&
  et~al.}]{2014ApJ...794..172A}
{Aab} A. {et~al.}, 2014, \apj, 794, 172

\bibitem[{{Aab} {et~al}\mbox{.}(2017){Aab}, {Abreu}, {Aglietta}, {Samarai},
  {Albuquerque}, {Allekotte}, {Almela}, {Alvarez Castillo},
  {Alvarez-Mu{\~n}iz}, {Anastasi}, \& et~al.}]{2017JCAP...06..026A}
{Aab} A. {et~al.}, 2017, \jcap, 6, 026

\bibitem[{Abreu {et~al}\mbox{.}(2013)Abreu {et~al.}}]{Abreu:2013kif}
Abreu P., {et~al.}, 2013, JCAP, 1305, 009

\bibitem[{Ade {et~al}\mbox{.}(2015)Ade {et~al.}}]{PLANCK2015}
Ade P. A.~R., {et~al.}, 2015

\bibitem[{{Allard}(2012)}]{2012APh....39...33A}
{Allard} D., 2012, Astroparticle Physics, 39, 33

\bibitem[{{Aloisio} \& {Berezinsky}(2004)}]{2004ApJ...612..900A}
{Aloisio} R., {Berezinsky} V., 2004, \apj, 612, 900

\bibitem[{Alves~Batista {et~al}\mbox{.}(2017)Alves~Batista, Shin, Devriendt,
  Semikoz, \& Sigl}]{AlvesBatista:2017vob}
Alves~Batista R., Shin M.-S., Devriendt J., Semikoz D., Sigl G., 2017

\bibitem[{{Araya-Melo} {et~al}\mbox{.}(2012){Araya-Melo}, {Arag{\'o}n-Calvo},
  {Br{\"u}ggen}, \& {Hoeft}}]{2012MNRAS.423.2325A}
{Araya-Melo} P.~A., {Arag{\'o}n-Calvo} M.~A., {Br{\"u}ggen} M., {Hoeft} M.,
  2012, \mnras, 423, 2325

\bibitem[{Armengaud {et~al}\mbox{.}(2007)Armengaud, Sigl, Beau, \&
  Miniati}]{CRPropa2006}
Armengaud E., Sigl G., Beau T., Miniati F., 2007, Astropart. Phys., 28, 463

\bibitem[{Armengaud {et~al}\mbox{.}(2005)Armengaud, Sigl, \&
  Miniati}]{finite_observer}
Armengaud E., Sigl G., Miniati F., 2005, Phys. Rev., D72, 043009

\bibitem[{Batista {et~al}\mbox{.}(2016)Batista, Dundovic, Erdmann, Kampert,
  Kuempel, M{\"u}ller, Sigl, van Vliet, Walz, \& Winchen}]{CRPropa2016}
Batista R.~A. {et~al.}, 2016, Journal of Cosmology and Astroparticle Physics,
  2016, 038

\bibitem[{{Beck}(2016)}]{2016A&ARv..24....4B}
{Beck} R., 2016, \aapr, 24, 4

\bibitem[{{Bernet} {et~al}\mbox{.}(2013){Bernet}, {Miniati}, \&
  {Lilly}}]{2013ApJ...772L..28B}
{Bernet} M.~L., {Miniati} F., {Lilly} S.~J., 2013, \apjl, 772, L28

\bibitem[{{Bertschinger}(1987)}]{1987ApJ...323L.103B}
{Bertschinger} E., 1987, \apjl, 323, L103

\bibitem[{{Bistolas} \& {Hoffman}(1998)}]{1998ApJ...492..439B}
{Bistolas} V., {Hoffman} Y., 1998, \apj, 492, 439

\bibitem[{{Blasi}(2013)}]{blasi}
{Blasi} P., 2013, \aapr, 21, 70

\bibitem[{Bonafede {et~al}\mbox{.}(2013)Bonafede, Vazza, Br{\"u}ggen, Murgia,
  Govoni, Feretti, Giovannini, \& Ogrean}]{Bonafede2013}
Bonafede A., Vazza F., Br{\"u}ggen M., Murgia M., Govoni F., Feretti L.,
  Giovannini G., Ogrean G., 2013, Mon. Not. Roy. Astron. Soc., 433, 3208

\bibitem[{{Broderick} {et~al}\mbox{.}(2012){Broderick}, {Chang}, \&
  {Pfrommer}}]{2012ApJ...752...22B}
{Broderick} A.~E., {Chang} P., {Pfrommer} C., 2012, \apj, 752, 22

\bibitem[{{Brown} {et~al}\mbox{.}(2017){Brown}, {Vernstrom}, {Carretti},
  {Dolag}, {Gaensler}, {Staveley-Smith}, {Bernardi}, {Haverkorn}, {Kesteven},
  \& {Poppi}}]{2017MNRAS.468.4246B}
{Brown} S. {et~al.}, 2017, \mnras, 468, 4246

\bibitem[{{Brown}(2011)}]{2011JApA...32..577B}
{Brown} S.~D., 2011, Journal of Astrophysics and Astronomy, 32, 577

\bibitem[{{Br{\"u}ggen} {et~al}\mbox{.}(2005){Br{\"u}ggen}, {Ruszkowski},
  {Simionescu}, {Hoeft}, \& {Dalla Vecchia}}]{2005ApJ...631L..21B}
{Br{\"u}ggen} M., {Ruszkowski} M., {Simionescu} A., {Hoeft} M., {Dalla Vecchia}
  C., 2005, \apjl, 631, L21

\bibitem[{{Bryan} {et~al}\mbox{.}(2014){Bryan}, {Norman}, {O'Shea}, {Abel},
  {Wise}, {Turk}, {Reynolds}, {Collins}, {Wang}, {Skillman}, {Smith},
  {Harkness}, {Bordner}, {Kim}, {Kuhlen}, {Xu}, {Goldbaum}, {Hummels},
  {Kritsuk}, {Tasker}, {Skory}, {Simpson}, {Hahn}, {Oishi}, {So}, {Zhao},
  {Cen}, {Li}, \& {Enzo Collaboration}}]{ENZO}
{Bryan} G.~L. {et~al.}, 2014, \apjs, 211, 19

\bibitem[{{Campbell}(2015)}]{2015MNRAS.448.2854C}
{Campbell} S.~S., 2015, \mnras, 448, 2854

\bibitem[{Das {et~al}\mbox{.}(2008)Das, Kang, Ryu, \& Cho}]{Das:2008vb}
Das S., Kang H., Ryu D., Cho J., 2008, Astrophys. J., 682, 29

\bibitem[{{Dawson} {et~al}\mbox{.}(2017){Dawson}, {Fukushima}, \&
  {Sokolsky}}]{2017arXiv170307897D}
{Dawson} B.~R., {Fukushima} M., {Sokolsky} P., 2017,
  arXiv:astro-ph.HE/1703.07897

\bibitem[{{Dedner} {et~al}\mbox{.}(2002){Dedner}, {Kemm}, {Kr{\"o}ner}, {Munz},
  {Schnitzer}, \& {Wesenberg}}]{Dedner2002}
{Dedner} A., {Kemm} F., {Kr{\"o}ner} D., {Munz} C.-D., {Schnitzer} T.,
  {Wesenberg} M., 2002, Journal of Computational Physics, 175, 645

\bibitem[{{di Matteo} \& {Tinyakov}(2017)}]{2017arXiv170602534D}
{di Matteo} A., {Tinyakov} P., 2017, arxiv:astro-ph.HE/1706.02534

\bibitem[{{Dolag}(2006)}]{2006AN....327..575D}
{Dolag} K., 2006, Astronomische Nachrichten, 327, 575

\bibitem[{{Dolag} {et~al}\mbox{.}(1999){Dolag}, {Bartelmann}, \&
  {Lesch}}]{1999dtrp.conf..237D}
{Dolag} K., {Bartelmann} M., {Lesch} H., 1999, in Diffuse Thermal and
  Relativistic Plasma in Galaxy Clusters, {Boehringer} H., {Feretti} L.,
  {Schuecker} P., eds., p. 237

\bibitem[{Dolag {et~al}\mbox{.}(2004)Dolag, Grasso, Springel, \&
  Tkachev}]{Dolag:2003ra}
Dolag K., Grasso D., Springel V., Tkachev I., 2004, JETP Lett., 79, 583, [Pisma
  Zh. Eksp. Teor. Fiz.79,719(2004)]

\bibitem[{Donnert {et~al}\mbox{.}(2009)Donnert, Dolag, Lesch, \&
  Muller}]{Donnert2008}
Donnert J., Dolag K., Lesch H., Muller E., 2009, Mon. Not. Roy. Astron. Soc.,
  392, 1008

\bibitem[{{Dova}(2016)}]{2016arXiv160407584D}
{Dova} M.~T., 2016, arxiv:astro-ph.HE/1604.07584

\bibitem[{Dundovi\'c \& Sigl(2017)}]{Dundovic:2017vsz}
Dundovi\'c A., Sigl G., 2017

\bibitem[{{Epele} \& {Roulet}(1998)}]{1998JHEP...10..009E}
{Epele} L.~N., {Roulet} E., 1998, Journal of High Energy Physics, 10, 009

\bibitem[{{Feretti} {et~al}\mbox{.}(2012){Feretti}, {Giovannini}, {Govoni}, \&
  {Murgia}}]{Feretti2012}
{Feretti} L., {Giovannini} G., {Govoni} F., {Murgia} M., 2012, \aapr, 20, 54

\bibitem[{{Ganon} \& {Hoffman}(1993)}]{1993ApJ...415L...5G}
{Ganon} G., {Hoffman} Y., 1993, \apjl, 415, L5

\bibitem[{{Hackstein} {et~al}\mbox{.}(2016){Hackstein}, {Vazza}, {Br{\"u}ggen},
  {Sigl}, \& {Dundovic}}]{Hackstein2016}
{Hackstein} S., {Vazza} F., {Br{\"u}ggen} M., {Sigl} G., {Dundovic} A., 2016,
  \mnras, 462, 3660

\bibitem[{Harari {et~al}\mbox{.}(2000)Harari, Mollerach, \&
  Roulet}]{Harari:2000az}
Harari D., Mollerach S., Roulet E., 2000, JHEP, 02, 035

\bibitem[{{Harari} {et~al}\mbox{.}(2002{\natexlab{a}}){Harari}, {Mollerach}, \&
  {Roulet}}]{2002JHEP...07..006H}
{Harari} D., {Mollerach} S., {Roulet} E., 2002{\natexlab{a}}, Journal of High
  Energy Physics, 7, 006

\bibitem[{{Harari} {et~al}\mbox{.}(2002{\natexlab{b}}){Harari}, {Mollerach},
  {Roulet}, \& {S{\'a}nchez}}]{2002JHEP...03..045H}
{Harari} D., {Mollerach} S., {Roulet} E., {S{\'a}nchez} F., 2002{\natexlab{b}},
  Journal of High Energy Physics, 3, 045

\bibitem[{{He{\ss}} {et~al}\mbox{.}(2013){He{\ss}}, {Kitaura}, \&
  {Gottl{\"o}ber}}]{2013MNRAS.435.2065H}
{He{\ss}} S., {Kitaura} F.-S., {Gottl{\"o}ber} S., 2013, \mnras, 435, 2065

\bibitem[{{Hillas}(1984)}]{hillas}
{Hillas} A.~M., 1984, Annual Review of Astronomy and Astrophysics, 22, 425

\bibitem[{{Hoffman} \& {Ribak}(1991)}]{1991ApJ...380L...5H}
{Hoffman} Y., {Ribak} E., 1991, \apjl, 380, L5

\bibitem[{{Hoffman} \& {Ribak}(1992)}]{1992ApJ...384..448H}
{Hoffman} Y., {Ribak} E., 1992, \apj, 384, 448

\bibitem[{{Ivanov}(2010)}]{2010ApJ...712..746I}
{Ivanov} A.~A., 2010, \apj, 712, 746

\bibitem[{{Jasche} \& {Wandelt}(2013)}]{2013MNRAS.432..894J}
{Jasche} J., {Wandelt} B.~D., 2013, \mnras, 432, 894

\bibitem[{{Kampert} {et~al}\mbox{.}(2013){Kampert}, {Kulbartz}, {Maccione},
  {Nierstenhoefer}, {Schiffer}, {Sigl}, \& {van Vliet}}]{CRPropa2013}
{Kampert} K.-H., {Kulbartz} J., {Maccione} L., {Nierstenhoefer} N., {Schiffer}
  P., {Sigl} G., {van Vliet} A.~R., 2013, Astroparticle Physics, 42, 41

\bibitem[{{Kim} {et~al}\mbox{.}(2016){Kim}, {Lilly}, {Miniati}, {Bernet},
  {Beck}, {O'Sullivan}, \& {Gaensler}}]{2016ApJ...829..133K}
{Kim} K.~S., {Lilly} S.~J., {Miniati} F., {Bernet} M.~L., {Beck} R.,
  {O'Sullivan} S.~P., {Gaensler} B.~M., 2016, \apj, 829, 133

\bibitem[{{Kitaura}(2013)}]{2013MNRAS.429L..84K}
{Kitaura} F.-S., 2013, \mnras, 429, L84

\bibitem[{{Klypin} {et~al}\mbox{.}(2003){Klypin}, {Hoffman}, {Kravtsov}, \&
  {Gottl{\"o}ber}}]{2003ApJ...596...19K}
{Klypin} A., {Hoffman} Y., {Kravtsov} A.~V., {Gottl{\"o}ber} S., 2003, \apj,
  596, 19

\bibitem[{Kotera \& Lemoine(2008)}]{Kotera:2007ca}
Kotera K., Lemoine M., 2008, Phys. Rev., D77, 023005

\bibitem[{{Kravtsov} {et~al}\mbox{.}(2002){Kravtsov}, {Klypin}, \&
  {Hoffman}}]{2002ApJ...571..563K}
{Kravtsov} A.~V., {Klypin} A., {Hoffman} Y., 2002, \apj, 571, 563

\bibitem[{Kurganov \& Tadmor(2000)}]{KurganovTadmor2000}
Kurganov A., Tadmor E., 2000, J. Comput. Phys., 160, 241

\bibitem[{{Lavaux}(2010)}]{2010MNRAS.406.1007L}
{Lavaux} G., 2010, \mnras, 406, 1007

\bibitem[{{Lavaux} {et~al}\mbox{.}(2008){Lavaux}, {Mohayaee}, {Colombi},
  {Tully}, {Bernardeau}, \& {Silk}}]{2008MNRAS.383.1292L}
{Lavaux} G., {Mohayaee} R., {Colombi} S., {Tully} R.~B., {Bernardeau} F.,
  {Silk} J., 2008, \mnras, 383, 1292

\bibitem[{{Lemoine} \& {Waxman}(2009)}]{2009JCAP...11..009L}
{Lemoine} M., {Waxman} E., 2009, \jcap, 11, 009

\bibitem[{{Letessier-Selvon}(2014)}]{2014BrJPh..44..560L}
{Letessier-Selvon} A., 2014, Brazilian Journal of Physics, 44, 560

\bibitem[{{Neronov} \& {Vovk}(2010)}]{NeronovVovk2010}
{Neronov} A., {Vovk} I., 2010, Science, 328, 73

\bibitem[{{Pierre Auger Collaboration}(2012)}]{2012ApJS..203...34P}
{Pierre Auger Collaboration}, 2012, \apjs, 203, 34

\bibitem[{{Planck Collaboration} {et~al}\mbox{.}(2014){Planck Collaboration},
  {Ade}, {Aghanim}, {Armitage-Caplan}, {Arnaud}, {Ashdown}, {Atrio-Barandela},
  {Aumont}, {Baccigalupi}, {Banday}, \& et~al.}]{2014A&A...571A..16P}
{Planck Collaboration} {et~al.}, 2014, \aap, 571, A16

\bibitem[{{Planck Collaboration} {et~al}\mbox{.}(2016){Planck Collaboration},
  {Ade}, {Aghanim}, {Arnaud}, {Arroja}, {Ashdown}, {Aumont}, {Baccigalupi},
  {Ballardini}, {Banday}, \& et~al.}]{2016A&A...594A..19P}
{Planck Collaboration} {et~al.}, 2016, \aap, 594, A19

\bibitem[{{Pshirkov} {et~al}\mbox{.}(2016){Pshirkov}, {Tinyakov}, \&
  {Urban}}]{2016PhRvL.116s1302P}
{Pshirkov} M.~S., {Tinyakov} P.~G., {Urban} F.~R., 2016, Physical Review
  Letters, 116, 191302

\bibitem[{{Ryu} {et~al}\mbox{.}(2012){Ryu}, {Schleicher}, {Treumann}, {Tsagas},
  \& {Widrow}}]{2012SSRv..166....1R}
{Ryu} D., {Schleicher} D.~R.~G., {Treumann} R.~A., {Tsagas} C.~G., {Widrow}
  L.~M., 2012, \ssr, 166, 1

\bibitem[{{Shu} \& {Osher}(1988)}]{ShuOsher1988}
{Shu} C.-W., {Osher} S., 1988, Journal of Computational Physics, 77, 439

\bibitem[{Sigl {et~al}\mbox{.}(2004)Sigl, Miniati, \& Ensslin}]{Sigl2004}
Sigl G., Miniati F., Ensslin T., 2004, Nucl. Phys. Proc. Suppl., 136, 224,
  [,224(2004)]

\bibitem[{Sigl {et~al}\mbox{.}(2003)Sigl, Miniati, \& Ensslin}]{Sigl:2003ay}
Sigl G., Miniati F., Ensslin T.~A., 2003, Phys. Rev., D68, 043002

\bibitem[{{Sigl} {et~al}\mbox{.}(2004){Sigl}, {Miniati}, \&
  {En{\ss}lin}}]{2004PhRvD..70d3007S}
{Sigl} G., {Miniati} F., {En{\ss}lin} T.~A., 2004, \prd, 70, 043007

\bibitem[{{Sorce}(2015)}]{2015MNRAS.450.2644S}
{Sorce} J.~G., 2015, \mnras, 450, 2644

\bibitem[{{Sorce} {et~al}\mbox{.}(2014){Sorce}, {Courtois}, {Gottl{\"o}ber},
  {Hoffman}, \& {Tully}}]{2014MNRAS.437.3586S}
{Sorce} J.~G., {Courtois} H.~M., {Gottl{\"o}ber} S., {Hoffman} Y., {Tully}
  R.~B., 2014, \mnras, 437, 3586

\bibitem[{{Sorce} {et~al}\mbox{.}(2016){Sorce}, {Gottl{\"o}ber}, {Yepes},
  {Hoffman}, {Courtois}, {Steinmetz}, {Tully}, {Pomar{\`e}de}, \&
  {Carlesi}}]{2016MNRAS.455.2078S}
{Sorce} J.~G. {et~al.}, 2016, \mnras, 455, 2078

\bibitem[{Stanev(1997)}]{0004-637X-479-1-290}
Stanev T., 1997, The Astrophysical Journal, 479, 290

\bibitem[{Takami {et~al}\mbox{.}(2012)Takami, Inoue, \&
  Yamamoto}]{Takami2012767}
Takami H., Inoue S., Yamamoto T., 2012, Astroparticle Physics, 35, 767

\bibitem[{Takami \& Sato(2008)}]{Takami:2007kq}
Takami H., Sato K., 2008, Astrophys. J., 681, 1279

\bibitem[{{The Pierre Auger Collaboration} {et~al}\mbox{.}(2017){The Pierre
  Auger Collaboration}, {Aab}, {Abreu}, {Aglietta}, {Samarai}, {Albuquerque},
  {Allekotte}, {Almela}, {Alvarez Castillo}, {Alvarez-Mu{\~n}iz}, \&
  et~al.}]{2017arXiv170907321T}
{The Pierre Auger Collaboration} {et~al.}, 2017, ArXiv astro-ph.HE/1709.07321

\bibitem[{Tinyakov \& Tkachev(2005)}]{Tinyakov:2004pw}
Tinyakov P.~G., Tkachev I.~I., 2005, Astropart. Phys., 24, 32

\bibitem[{{Tinyakov} \& {Urban}(2015)}]{2015JETP..120..533T}
{Tinyakov} P.~G., {Urban} F.~R., 2015, Soviet Journal of Experimental and
  Theoretical Physics, 120, 533

\bibitem[{{Trivedi} {et~al}\mbox{.}(2014){Trivedi}, {Subramanian}, \&
  {Seshadri}}]{2014PhRvD..89d3523T}
{Trivedi} P., {Subramanian} K., {Seshadri} T.~R., 2014, \prd, 89, 043523

\bibitem[{{Tully} {et~al}\mbox{.}(2013){Tully}, {Courtois}, {Dolphin},
  {Fisher}, {H{\'e}raudeau}, {Jacobs}, {Karachentsev}, {Makarov}, {Makarova},
  {Mitronova}, {Rizzi}, {Shaya}, {Sorce}, \& {Wu}}]{2013AJ....146...86T}
{Tully} R.~B. {et~al.}, 2013, \aj, 146, 86

\bibitem[{Vall{\'e}e(2004)}]{galactic_fields}
Vall{\'e}e J.~P., 2004, New Astronomy Reviews, 48, 763

\bibitem[{{van de Weygaert} \& {Bertschinger}(1996)}]{1996MNRAS.281...84V}
{van de Weygaert} R., {Bertschinger} E., 1996, \mnras, 281, 84

\bibitem[{van Vliet(2014)}]{AvanVlietPhD}
van Vliet A.~R., 2014, PhD thesis, Hamburg University

\bibitem[{{Vazza} {et~al}\mbox{.}(2011){Vazza}, {Dolag}, {Ryu}, {Brunetti},
  {Gheller}, {Kang}, \& {Pfrommer}}]{2011MNRAS.418..960V}
{Vazza} F., {Dolag} K., {Ryu} D., {Brunetti} G., {Gheller} C., {Kang} H.,
  {Pfrommer} C., 2011, \mnras, 418, 960

\bibitem[{Vazza {et~al}\mbox{.}(2015)Vazza, Ferrari, Br{\"u}ggen, Bonafede,
  Gheller, \& Wang}]{Vazza2015}
Vazza F., Ferrari C., Br{\"u}ggen M., Bonafede A., Gheller C., Wang P., 2015,
  Astron. Astrophys., 580, A119

\bibitem[{{Vernstrom} {et~al}\mbox{.}(2017){Vernstrom}, {Gaensler}, {Brown},
  {Lenc}, \& {Norris}}]{2017MNRAS.467.4914V}
{Vernstrom} T., {Gaensler} B.~M., {Brown} S., {Lenc} E., {Norris} R.~P., 2017,
  \mnras, 467, 4914

\bibitem[{{Wang} {et~al}\mbox{.}(2014){Wang}, {Mo}, {Yang}, {Jing}, \&
  {Lin}}]{2014ApJ...794...94W}
{Wang} H., {Mo} H.~J., {Yang} X., {Jing} Y.~P., {Lin} W.~P., 2014, \apj, 794,
  94

\bibitem[{{Wang} {et~al}\mbox{.}(2010){Wang}, {Abel}, \&
  {Kaehler}}]{2010NewA...15..581W}
{Wang} P., {Abel} T., {Kaehler} R., 2010, \na, 15, 581

\bibitem[{Yoshiguchi {et~al}\mbox{.}(2003)Yoshiguchi, Nagataki, Tsubaki, \&
  Sato}]{Yoshiguchi:2002rb}
Yoshiguchi H., Nagataki S., Tsubaki S., Sato K., 2003, Astrophys. J., 586,
  1211, [Erratum: Astrophys. J.601,592(2004)]

\end{thebibliography}
